\documentclass[twocolumn,showpacs,secnumarabic,preprintnumbers,amsmath,amssymb,10pt,a4paper,superscriptaddress,aps,pre,showpacs]{revtex4-1}

\usepackage{amsmath}
\usepackage{amsfonts}
\usepackage{amssymb}

\usepackage{todonotes}
\usepackage{graphicx}
\usepackage{verbatim}
\usepackage{color}
\usepackage{hyperref}

\newcommand*{\eql}[1]{\label{eqn:#1}}
\newcommand*{\eqr}[1]{\eqref{eqn:#1}}

\newcommand*{\figl}[1]{\label{fig:#1}}
\newcommand*{\figr}[1]{\ref{fig:#1}}

\newcommand*{\secl}[1]{\label{sec:#1}}
\newcommand*{\secr}[1]{\ref{sec:#1}}

\newcommand*{\half}{\frac{1}{2}}
\renewcommand*{\i}{\mathrm{i}}
\newcommand*{\e}{\mathrm{e}}

\newcommand*{\dd}{\mathrm{d}}   
\newcommand*{\DD}{\mathrm{D}}   
\newcommand*{\CD}{\mathcal{D}}  

\newcommand*{\inti}[1]{\int \! \dd #1 \,}               
\newcommand*{\intlu}[3]{\int_{#2}^{#3} \! \dd #1 \,}    
\newcommand*{\intDlu}[3]{\int_{#2}^{#3} \! \DD #1 \,}   
\newcommand*{\intCi}[1]{\int \! \CD #1 \,}              

\newcommand*{\pdiff}[2]{\frac{\partial #1}{\partial #2}}                
\newcommand*{\adiff}[2]{\frac{\delta #1}{\delta #2}}                    
\newcommand*{\madiff}[3]{\frac{\delta^2 #1}{\delta #2 \delta #3}}       

\newcommand*{\plr}[1]{\left(#1\right)}

\newcommand*{\Blr}[1]{\left\{#1\right\}}
\newcommand*{\tlr}[1]{\left\langle#1\right\rangle}
\newcommand*{\mlr}[1]{\left|#1\right|}

\newcommand*{\mean}[1]{\tlr{#1}}
\newcommand*{\eval}[2]{\left. #1 \right|_{#2}}  

\newcommand*{\Oh}[1]{\mathcal{O}\plr{#1}}

\newenvironment{aligncn}{\begin{equation}\begin{aligned}}{\end{aligned}\end{equation}}

\def\clap#1{\hbox to 0pt{\hss#1\hss}}
\def\mathclap{\mathpalette\mathclapinternal}
\def\mathclapinternal#1#2{%
\clap{$\mathsurround=0pt#1{#2}$}}

\newcommand*{\htx}{\widehat{x}}
\newcommand*{\htC}{\widehat{C}}
\newcommand*{\htK}{\widehat{K}}
\newcommand*{\htL}{\widehat{L}}
\newcommand*{\hteta}{\widehat{\eta}}

\newcommand*{\tix}{\widetilde{x}}
\newcommand*{\tiH}{\widetilde{H}}
\newcommand*{\tieta}{\widetilde{\eta}}
\newcommand*{\tixi}{\widetilde{\xi}}

\newcommand{\otherplayers}[1]{#1+1, \ldots, #1-1}
\newcommand{\otheractions}[2]{#2_{#1+1}, \ldots, #2_{#1-1}}
\newcommand{\allactions}[2]{#1_1, \ldots, #1_{#2}}
\newcommand{\otherstrategies}[3]{\prod_{\kappa \neq #1} x^\kappa_{#2_\kappa}(#3)}
\newcommand{\otherstrategiesalt}[4]{\prod_{#4 \neq #1} x^{#4}_{#2_{#4}}(#3)}
\newcommand{\pdimpayoffs}[2]{\Pi^{#1}_{#2, \otheractions{#1}{#2}}}
\newcommand{\pdimpayoffsalt}[3]{\Pi^{#1}_{#3, \otheractions{#1}{#2}}}

\newcommand{\eff}{\textrm{eff}}

\usepackage{bm}
\newcommand{\be}{\begin{equation}}
\newcommand{\ee}{\end{equation}}
\newcommand{\bd}{\begin{displaymath}}
\newcommand{\ed}{\end{displaymath}}
\newcommand{\BE}{\begin{eqnarray}}
\newcommand{\EE}{\end{eqnarray}}

\newcommand{\erf}{{\rm erf}}

\newcommand{\bx}{\ensuremath{\mathbf{x}}}

\begin{document}
\title{The prevalence of chaotic dynamics in games with many players}

\author {James BT Sanders}
\email{james.sanders-2@postgrad.manchester.ac.uk }
\affiliation{Theoretical Physics, School of Physics and Astronomy, The University of Manchester, Manchester M13 9PL, UK}

\author{Tobias Galla}
\email{tobias.galla@manchester.ac.uk}
\affiliation{Theoretical Physics, School of Physics and Astronomy, The University of Manchester, Manchester M13 9PL, UK}

\author{J. Doyne Farmer}
\email{doyne.farmer@oxfordmartin.ox.ac.uk}
\affiliation{Institute for New Economic Thinking at the Oxford Martin School, University of Oxford, Oxford OX2 6ED, UK}
\affiliation{Mathematical Institute, University of Oxford, Oxford OX1 3LP, UK}
\affiliation{Santa Fe Institute, Santa Fe, NM 87501, UK}
\begin{abstract}
We study adaptive learning in a typical $p$-player game.   The payoffs of the games are randomly generated and then held fixed.  The strategies of the players evolve through time as the players learn.   The trajectories in the strategy space display a range of qualitatively different behaviors, with attractors that include unique fixed points, multiple fixed points, limit cycles and chaos.  In the limit where the game is complicated, in the sense that the players can take many possible actions, we use a generating-functional approach to establish the parameter range in which learning dynamics converge to a stable fixed point.  The size of this region goes to zero as the number of players goes to infinity, suggesting that complex non-equilibrium behavior, exemplified by chaos, may be the norm for complicated games with many players.
\end{abstract}

\maketitle
\section{Introduction}
Many branches of science are interested in systems made up of a large number of competing individuals.  Examples range from financial markets and other social systems, to populations undergoing biological evolution, to networked computer systems.  In many such situations individuals compete for limited resources, and the natural model  is a game, which consists of a set of players who at any point in time choose from a set of possible actions in an attempt to maximize their payoff.  Game theory has received a great deal of attention since the mid-20th century \cite{vonneumann}, but research has overwhelmingly focused on simple games, with only a very small number of players and pure strategies, even though real game-like systems often involve large numbers of individuals and possible strategies.  While many of the observed properties of simple games carry over directly to more complicated ones, it is becoming increasingly clear that complicated games can show important types of behavior not found in simpler systems.

Early work in game theory focused on the concept of equilibria, in particular the famous {\em Nash equilibrium} \cite{nash}, in which the players adopt strategies such that no player can improve her own payoff by unilaterally changing her own strategy. The strategies at a Nash equilibrium can be probabilistic combinations of pure strategies, called {\em mixed strategies}. Nash's ideas have been particularly influential in economics, where agents are often assumed to adopt Nash equilibria.  It should be emphasized that this is a behavioral assumption, and that the empirical evidence for this is mixed \cite{Farmer2008}.

Equilibrium models are perfectly plausible in the context of simple games when there is a unique Nash equilibrium that is easy to calculate.  In complicated games, there are typically numerous distinct Nash equilibria~\cite{berg, berg2}, and locating even one of them can be a laborious task: the computing time of the best known algorithms increases exponentially with the size of the game \cite{Daskalakis}. This seems to cast doubt on whether it is reasonable to assume that players of complicated games naturally discover equilibria. But if they don't play equilibria, what do they do instead?  And is there a way to determine {\it a priori} whether players will converge to a unique Nash equilibrium using a reasonable learning strategy?

Opper and Diederich studied replicator dynamics, a standard model of biological evolution, in the context of complicated games.  They found that the dynamics converges to a unique fixed point in some regions of parameter space, but that in other regions, the dynamics does not settle down~\cite{opper,opper2}.  Sato and co-workers showed that adaptation learning can result in chaotic dynamics even in low-dimensional games \cite{sato,sato2,sato3}.  Building on this earlier work, Galla and Farmer studied complicated two-player games in which the strategies are randomly generated but fixed in time, assuming that players use an experience-weighted attraction dynamics to learn their strategies \cite{GallaFarmer}.  They found that there are distinct regions of the parameter space with different behaviors.  When the timescale for learning is short and the payoffs of the players are strongly negatively correlated, they observed convergence to unique fixed points.  But when the time scale for learning is long and when the payoffs are less negatively correlated they observed limit cycles and chaos.  And when the timescale for learning is long and the payoffs are positively correlated they observed a large multiplicity of stable fixed points.

In this paper we extend the work of \cite{GallaFarmer} to $p$-player games and find that, for large numbers of players, complex dynamics is not merely frequent but ubiquitous.  The region of parameter space in which the players' strategies consistently converge to a unique fixed point appears to vanish as $p \rightarrow \infty$.  This suggests that complex non-equilibrium behavior may be the norm in dynamics on complicated games with many players, at least under the type of learning we study here.

Our goal can be understood through an analogy to fluid flow.  It is well known that fluid flow can be characterized {\it a priori} in terms of a few key parameters that can be estimated on the back of an envelope.  The most famous of these is the Reynolds number, which is a non-dimensional ratio of stress and viscosity.  Though the precise transition point depends on other parameters, if the Reynolds number is high then the flow is likely to be turbulent.  Our goal here is similar:  We seek parameters that can characterize {\it a priori} whether or not a game will exhibit complex dynamics in the strategy space as the players learn.  Here we are particularly interested in what happens as the number of players increases.  Since the presence of many players makes the game more complex, we hypothesize that it will tend to make the strategy dynamics more complex as well.  (And indeed this is what we observe).

The remainder of the paper is organized as follows: In Sec. \ref{sec:model} we introduce the experience-weighted attraction learning algorithm and we define what we mean by a complex $p$-player game. Sec. \ref{sec:behavior} then contains an overview of the different types of behavior seen in the learning of such games, along with a more quantitive analysis based on numerical simulations. In Sec. \ref{sec:gfa} we then turn to a semi-analytical study of $p$-player learning based on a generating-functional approach. This technique allows us to derive estimates for the boundaries of stable and complex behavior in parameter space for games with a large number of strategies. We then turn to a brief discussion of volatility clustering and the relevance of our result for the modeling of financial markets in Sec. \ref{sec:vol}, before we summarize our results. The Appendix contains further details of the numerical methods used to identify the different types of dynamical behavior, and of the generating functional analysis. It also contains some additional numerical results.

\section{Model}\label{sec:model}

\subsection{Experience-weighted attraction}
Suppose that a set of players repeatedly play a game in which they each choose from a distinct set of strategies, without conferring with each other. The players have good memories and a full understanding of the payoffs that a given combination of strategies would yield for them, and are only interested in maximizing their own payoffs.   We are interested in the case where the players learn their strategies based on past experience.  A common approach assumes the players adapt behavior over time based on the past success of each possible strategy \cite{ho,camerer1,camerer,fudenberg}.  The basic idea is that the players calculate a numerical score, known as an `attraction' or `propensity', for each possible strategy, describing how successful they expect it to be. They then select a strategy based on the relative score of each possible action, play the game one or more times, then use the outcome to update their attractions for future play. This defines an adaptation process, in which agents learn from past experience and continuously try to improve their actions.

Two types of simple learning models have proved especially popular over the years.  In reinforcement learning, the players calculate the attraction of a strategy based on how successful it has been when they have employed it in the past.  In belief-based learning, the attractions are determined according to how successful the possible strategies would have been, had they been used in prior iterations.  The experience-weighted attraction (EWA) system, introduced by Camerer and Ho \cite{ho}, combines these two approaches into a single algorithm---in fact, belief-based learning can be seen as a deterministic limit of reinforcement learning in which the players sample all pure strategies at each time step.  Combined with a logit model of how the players choose strategies based on their attractions, EWA is observed to be a reasonably good match for how people learn to play simple games~\cite{camerer,camerer1}.

We are interested in how often EWA converges to equilibria, so we select the deterministic (belief-based) version of the model.  While the noisy (reinforcement) version may well perform stochastic oscillations about equilibria in the right conditions, we would expect that in general, the introduction of noise would lead to complex dynamics being observed even more often.  Both cases were studied by Galla and Farmer and the differences were not dramatic (see the Supplementary Material of \cite{GallaFarmer}).

Consider a $p$-player game, where each player has $N$ actions to choose from. The rewards for the players are defined by the generalised payoff matrix $\pdimpayoffs{\mu}{i}$, which represents the payoff to player~$\mu$ if they play action~$i$, while the other players $\otherplayers{\mu}$ play actions $\otheractions{\mu}{i}$, respectively (where the subscripts labelling the players are to be interpreted modulo~$p$). 

We use updates rules similar to those of \cite{GallaFarmer}, but adapted to the multi-player game,

\begin{aligncn} \eql{attraction_update}
  x^\mu_i(t+1) &= \frac{\exp[\beta Q^\mu_i(t)]}{\sum_k \exp[\beta Q^\mu_k(t)]}, \\
  Q^\mu_i(t+1) &= (1-\alpha) Q^\mu_i(t) + \sum_{\mathclap{\otheractions{\mu}{i}}} \pdimpayoffs{\mu}{i} \otherstrategies{\mu}{i}{t}.
\end{aligncn}

Here $\bx$ represents the players' strategies, with $x^\mu_i(t)$ denoting the probability that player~$\mu$ will choose action~$i$ at time~$t$.  The value $Q^\mu_i(t)$ is player~$\mu$'s {\it attraction} to action~$i$ at time~$t$.  The two parameters of the system are the memory loss rate $\alpha$, which lies in the interval $[0, 1]$, and the {\it intensity of choice} $\beta$, which is non-negative.  A player's attraction to an action is essentially a geometrically discounted average of the payoffs that would have been achieved by playing that action in earlier time steps, with a discount factor determined by $\alpha$.  The intensity of choice $\beta$ determines the bias with which players choose actions with higher attractions---if $\beta = 0$, then the players ignore the attractions and choose each action with equal probability, while in the limit as $\beta \rightarrow \infty$, the players each choose their most attractive action with probability~$1$.  The intensity of choice therefore plays a similar role to the inverse temperature in thermodynamics.

Our system is identical to that studied by Galla and Farmer in reference \cite{GallaFarmer}, except that they restricted the number of players to be  $p = 2$.  Our contribution is to understand how this changes as $p$ becomes larger.  The system of Camerer and Ho~\cite{camerer,camerer1}, as well as allowing for noise, allows the intensity of choice~$\beta$ to vary over time. In the present work we assume that it has attained its long-term value.  Thus the dynamics we study here are a special case of \cite{camerer,camerer1}.

The attractions $Q_i^\mu$ can be eliminated from the update rules in Eq. \eqr{attraction_update} to yield
\BE 
  x^\mu_i(t+1) &=& \frac{1}{Z^\mu(t)}x^\mu_i(t)^{1-\alpha}\nonumber \\
 && \hspace{-5em}\times \exp\plr{\beta \!\!\! \sum_{\otheractions{\mu}{i}} \!\!\! \pdimpayoffs{\mu}{i} \otherstrategies{\mu}{i}{t}},\label{eq:ewa_disc}
\EE
where
\BE
  Z^\mu(t) &=& \sum_k x^\mu_k(t)^{1-\alpha} \nonumber \\
  &&\exp\plr{\beta \!\!\! \sum_{\otheractions{\mu}{i}} \!\!\! \pdimpayoffsalt{\mu}{i}{k} \otherstrategies{\mu}{i}{t}}
\EE
is a normalization factor.  

Following~\cite{GallaFarmer}, we focus on the continuous-time limit of the EWA system. Letting
\begin{equation}
\nonumber
r = \beta/\alpha,
\end{equation}
this limit is found by keeping $r$ fixed while taking $\alpha\to 0$ and $\beta \to 0$ simultaneously, and rescaling time. Both the geometric discounting of the past profits of an action and the logit selection remain significant in this limit.  This yields the so-called Sato-Crutchfield dynamics \cite{sato,sato2}
\begin{equation} \eql{ewa_cts}
  \frac{\dot{x}^\mu_i(t)}{x^\mu_i(t)} = - \frac{1}{r} \ln x^\mu_i(t) + \sum_{\mathclap{\otheractions{\mu}{i}}} \pdimpayoffs{\mu}{i} \otherstrategies{\mu}{i}{t} - \rho^\mu(t),
\end{equation}
where $\rho^\mu(t) = \ln Z^\mu(t) / \beta$.  A detailed derivation can be found for example in the Supplementary Material of \cite{GallaFarmer}.  Note that, since each player satisfies the constraint that the probability of taking any given action sums to one, the resulting dynamical system for the learning dynamics is of dimension $(N-1)p$. 

We assume throughout this work that $\alpha$ and $\beta$ are small enough that the continuous limit is a good approximation. We take this limit mainly for analytical convenience; the continuous limit is easier to study, as it has only one relevant parameter, the ratio of $\alpha$ to $\beta$. 

When considering large values of~$N$ it is convenient to rescale the probabilities $x^\mu_i(t)$ so that they sum to $N$ for each player.  This means that $x^\mu_i(t) = \Oh{1}$, and so the objects inside the exponentials in Eq. (\ref{eq:ewa_disc}) remain finite as $N \rightarrow \infty$.  This can be achieved by modifying the definition of the normalization factors $Z^\mu$ accordingly. The payoff matrix entries are rescaled as well as explained below.

\subsection{Constructing typical complicated games with $p$ players}
We are interested in the behavior of the EWA system for generic complicated games.  To that end, we draw the payoff matrix ${\bf \Pi}$ from a multivariate normal distribution with
\begin{equation}
  \mean{\pdimpayoffsalt{\mu}{i}{i_\mu} \pdimpayoffsalt{\nu}{i}{i_\nu}} =
    \begin{cases}
      \frac{1}{N^{p-1}} & \mu = \nu \\
      \frac{\Gamma}{(p-1)N^{p-1}} & \mu \neq \nu
    \end{cases}\label{eq:payoffs}
\end{equation}
and all other correlations zero.  The multivariate Gaussian distribution is a natural choice because it is the maximum entropy distribution when there are constraints on the first and second moments.  We have chosen the construction above  because it yields the only possible multivariate Gaussian distribution that satisfies the following properties: (i) the distribution is symmetric with respect to the different players and pure strategies (that is, swapping any two players or pure strategies would leave the distribution unchanged) and (ii) the payoffs for any two distinct choices of actions are uncorrelated (by choice of actions we mean the tuple $(i_1,\dots,i_p)$ representing the pure strategies played by the $p$ players).  As before, we wish to emphasize that once $\Pi$ is chosen it remains fixed through the duration of the iterated game.  

The parameter $\Gamma$ can be seen as the level of `zero-sumness' of the game, and must lie on the interval $[-1, p-1]$.  When $\Gamma = p-1$, any outcome of the game leads to each player receiving the same payoff with probability 1.  When $\Gamma = 0$, all payoffs are uncorrelated.  When $\Gamma = -1$, for any given outcome, the players' payoffs sum to zero with probability 1.  Therefore $\Gamma$ can be seen as a `competition parameter'---at large positive values, the players have common goals, while at large negative values, they are working against each other.

When $p = 2$, the possible values of $\Gamma$  span the range $-1 \le \Gamma \le 1$.  However when $p > 2$ the situation becomes more complicated.  Eq. (\ref{eq:payoffs}) indicates that the payoffs each have variance $1/N^{p-1}$, the correlations between payoffs for different outcomes are uncorrelated, and the correlation between the payoffs of two different players for any given outcome is $\Gamma/((p-1)N^{p-1})$.  The scaling of the payoff matrix elements with $N^{-(p-1)}$ ensures that the objects inside the exponentials in Eq. (\ref{eq:ewa_disc}) remain finite as $N\to\infty$, so that both the memory-loss ($\alpha$) and payoff ($\beta$) factors remain significant in this limit. 

\begin{figure*}
  \centering
 
    \includegraphics[scale=0.5]{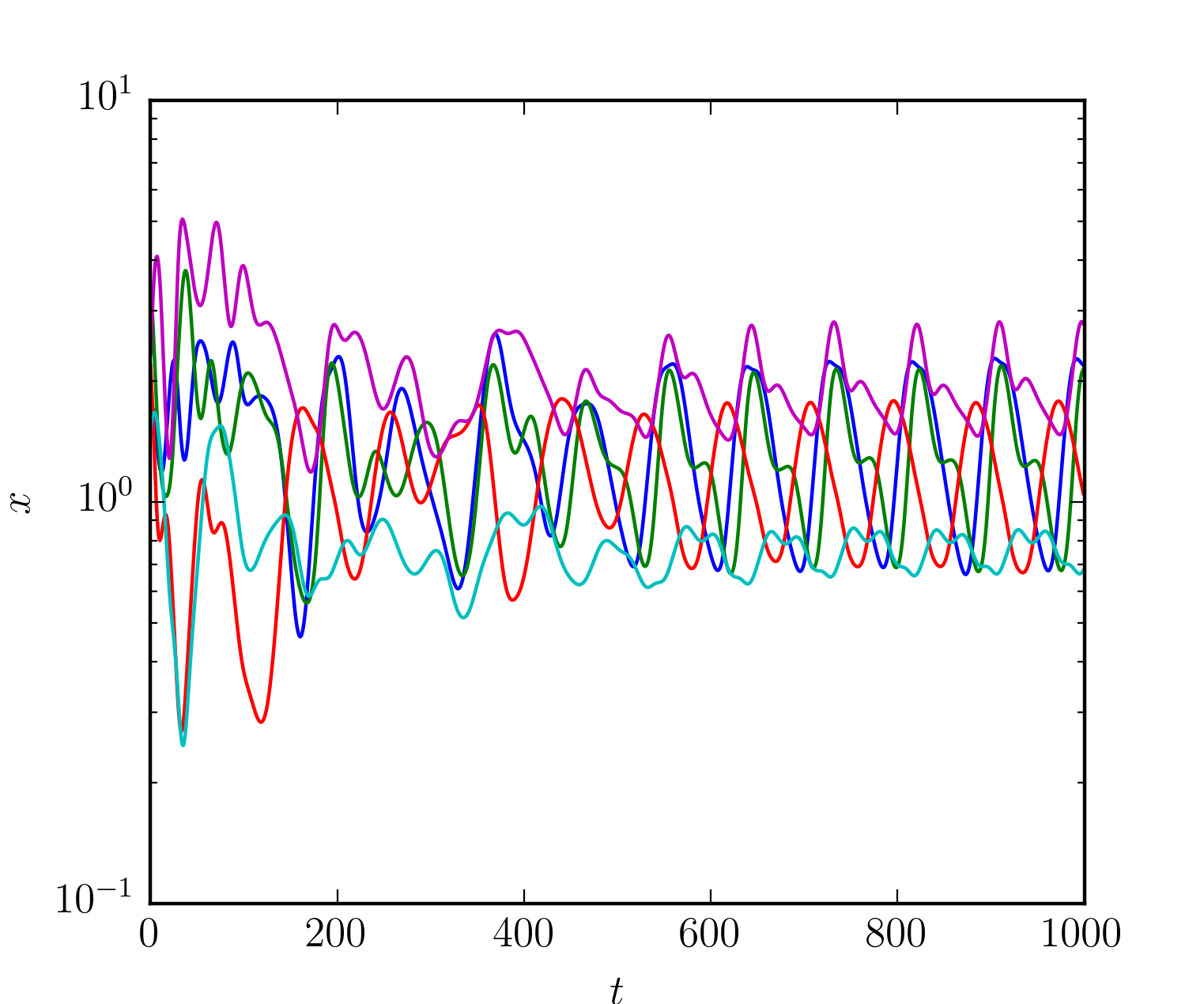}
    \includegraphics[scale=0.5]{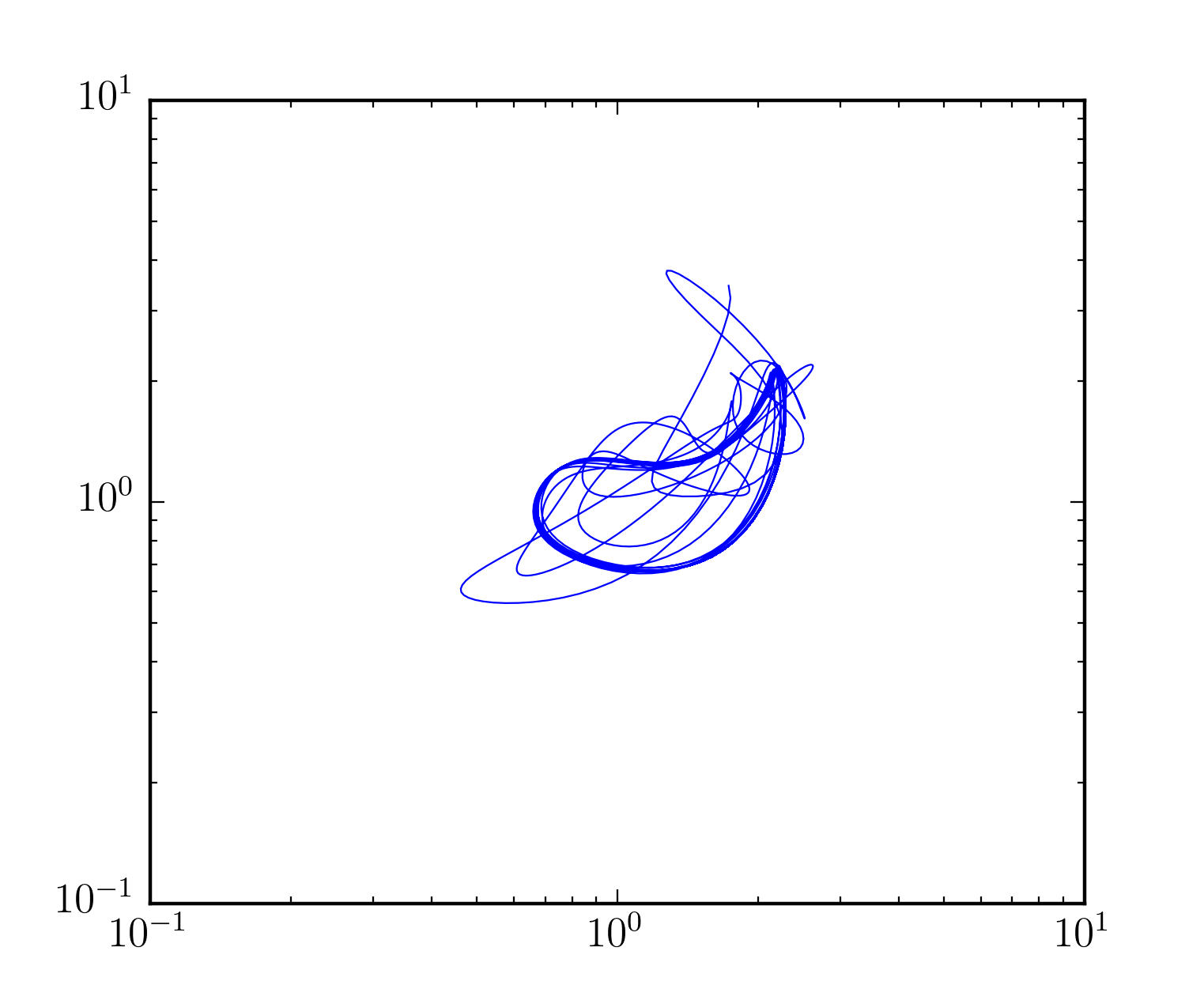}
    \\
 (a) Limit cycle ($\alpha = 0.038$).
  \\
    \includegraphics[scale=0.5]{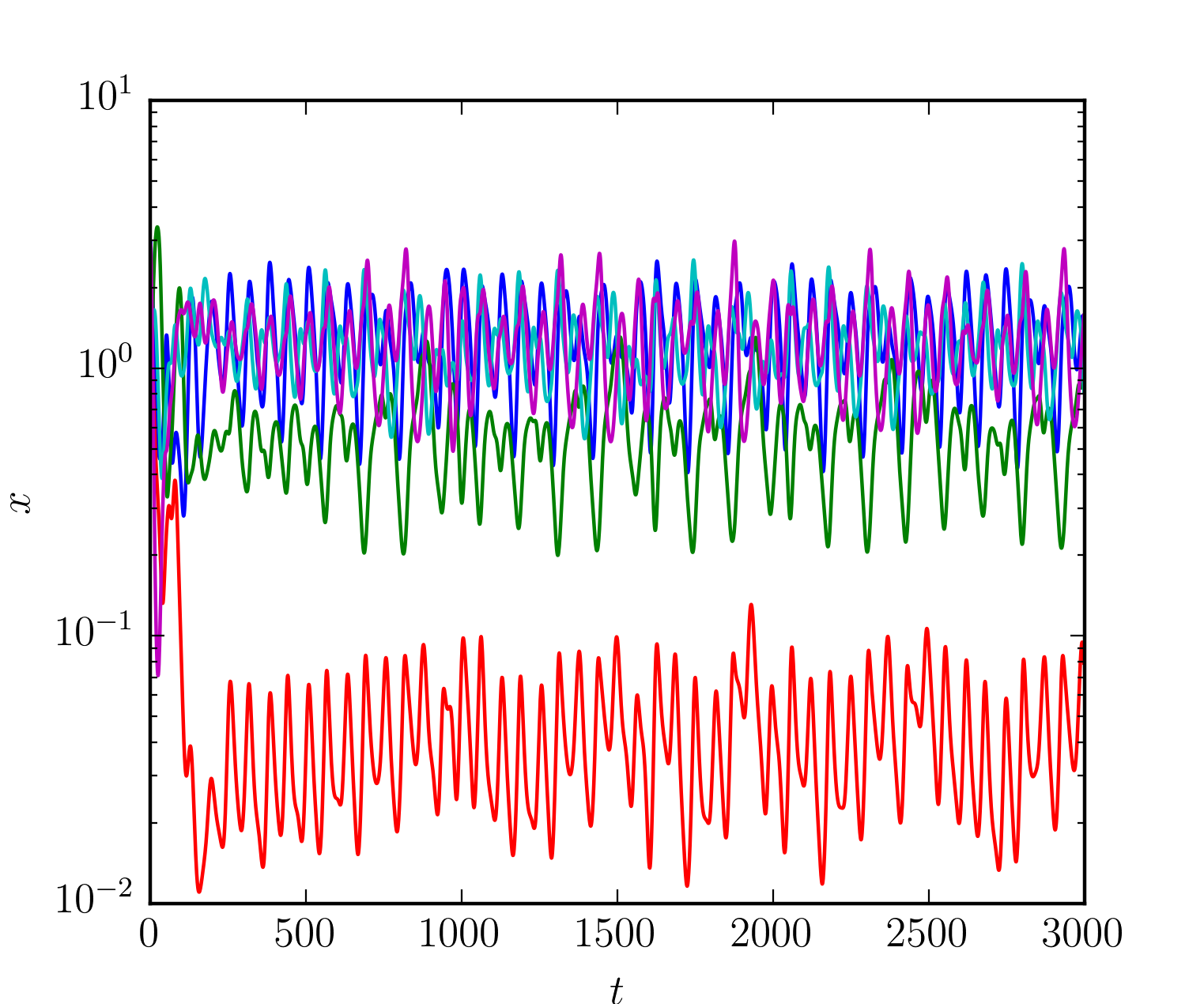}
    \includegraphics[scale=0.5]{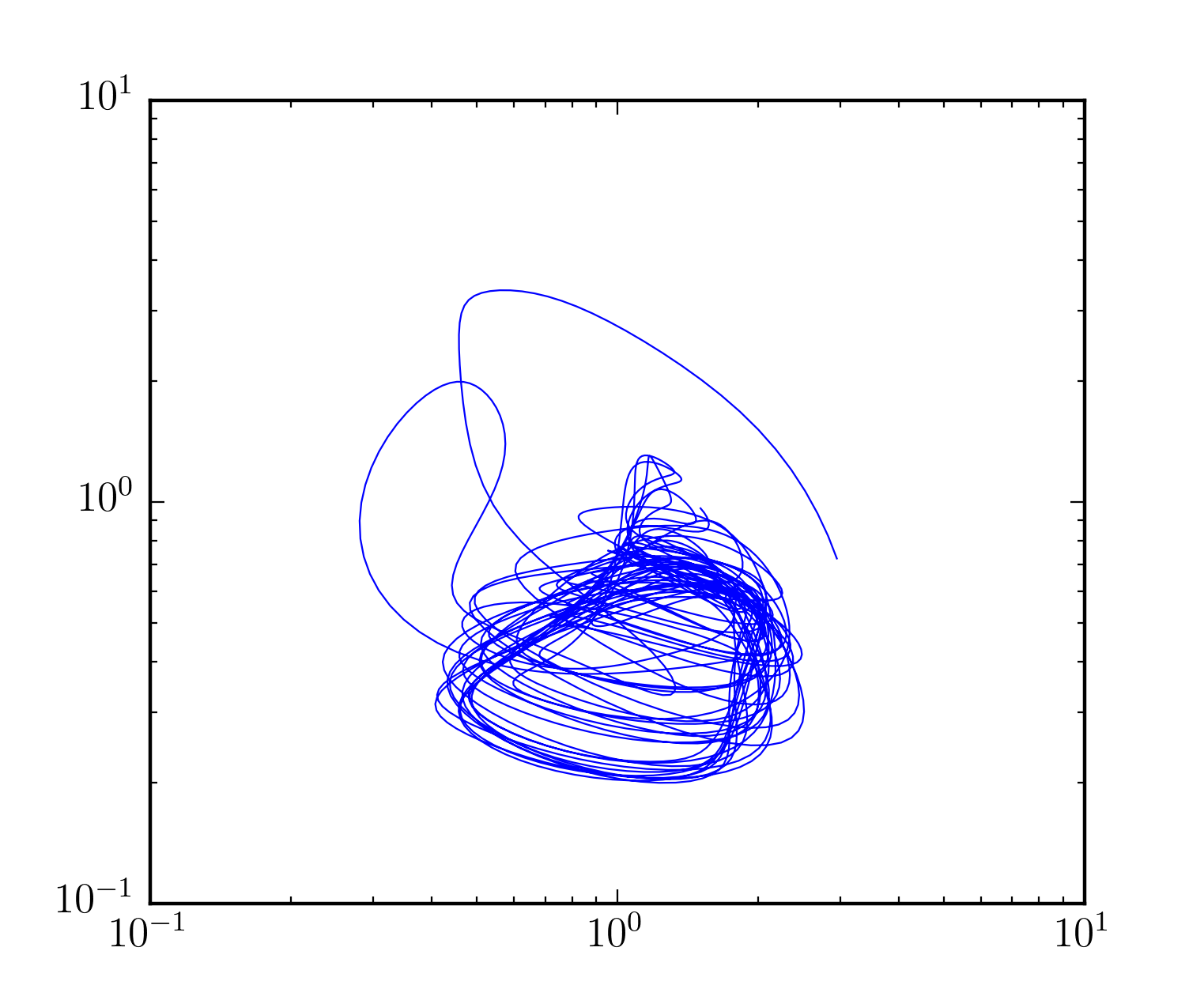}
    \\
    (b) Low-dimensional chaos ($\alpha = 0.037$).
    \\
   \includegraphics[scale=0.5]{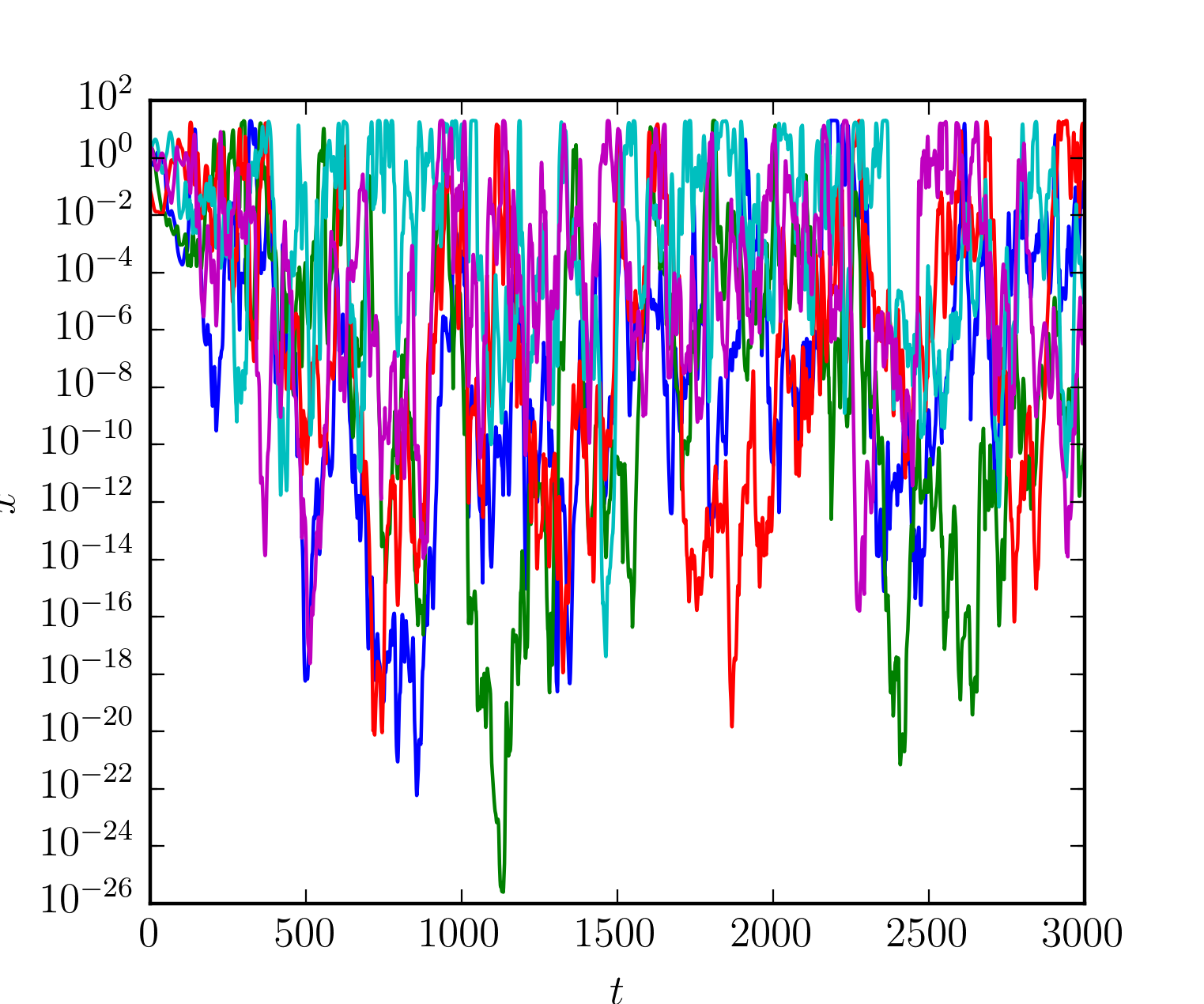}
    \includegraphics[scale=0.5]{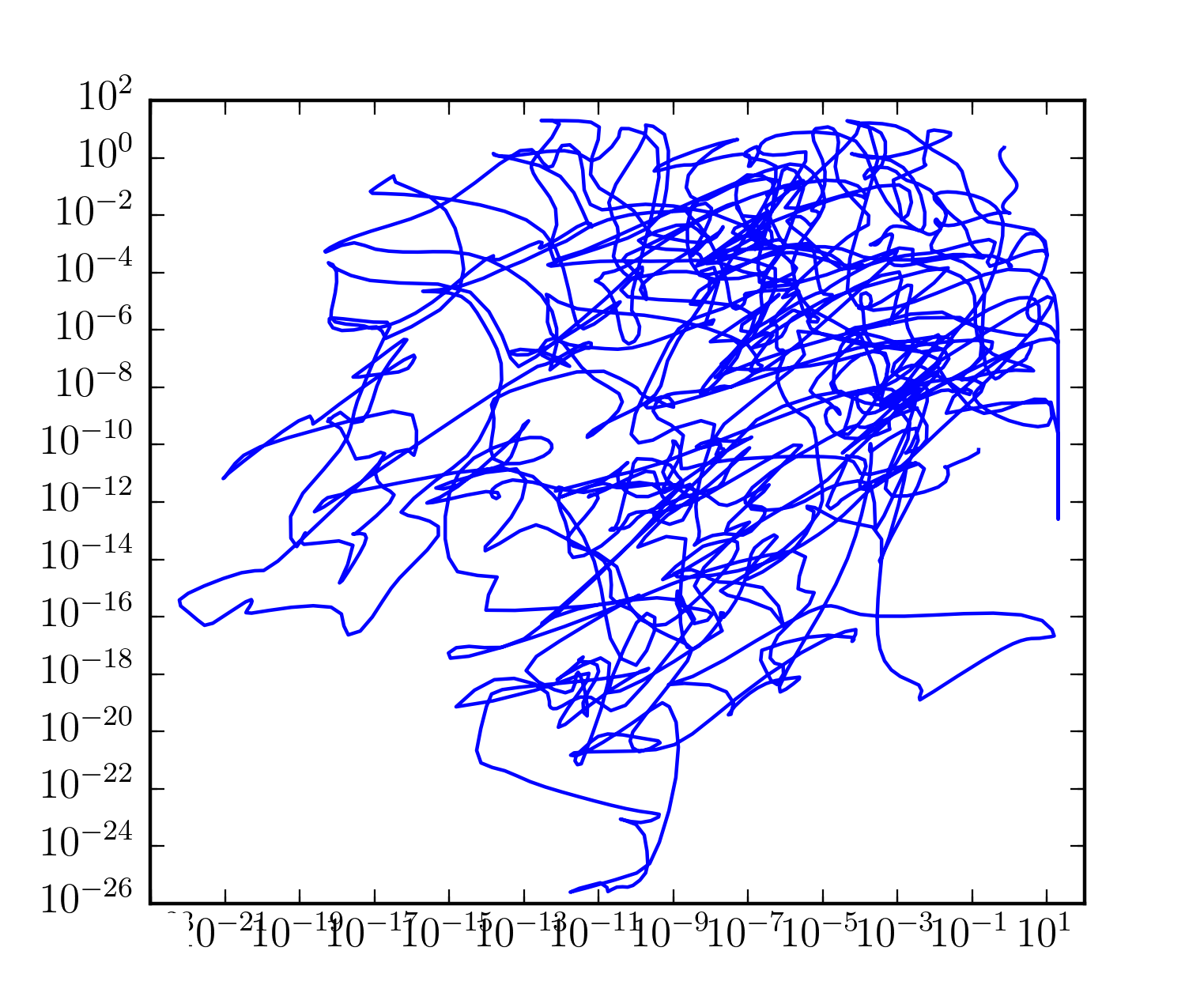}\\
    (c) High-dimensional chaos ($\alpha = 0.01$).
 
  \caption{Time series and phase plots showing complex dynamics under EWA learning, including (a) limit cycle, (b) low dimensional chaos, and (c) high dimensional chaos.   The game has three players ($p=3$) and $N= 20$ possible actions, with $\beta = 0.05$ and $\Gamma = -0.5$. The time series plots on the left show the probability $x_i^\mu$ for player $\mu$ to use action $i$ as a function of time for five different actions, and the phase plots on the right shows the probability for two of the actions as a function of each other.  Case (a) illustrates that limit cycles can have complicated geometric forms and long periods.  For smaller values of $\alpha/\beta$ and negative $\Gamma$, chaos is very common, ranging from low-dimensional chaotic attractors as shown in (b) to high dimensional attractors as shown in (c).  Note that for high dimensional chaos the probability that a given action is used at different points in time can vary by as much as a factor of $10^{20}$.}
  \figl{lc_chaos_time_series}
\end{figure*}
 
\subsection{Strategy for exploring the parameter space}

For a $p$-player game the payoff ``matrices" each have $p$ indices, and so are not two dimensional matrices in the normal usage of the world, but are $p$-dimensional.  This significantly complicates the problem of exploring the parameter space numerically.   For a game in which a single player can take one of $N$ actions the payoff matrix for a single player has $N^p$ components; for $p$ players there are $pN^p$ components.  For $p=10$ and $N=10$, for example, this means that $10^{11}$ (a hundred billion) random numbers must be generated in order to construct the game.  The sheer amount of memory needed for simulation creates a serious bottleneck.

Given the numerical constraints this forces us to rely more heavily on the analytic calculation than Galla and Farmer did when they explored two-player games.  We use numerical simulations to get a feeling for the behavior of the system, with relatively small values of $N$ and $p$. In parallel we perform an analytic calculation of the stability boundary between the region where there is unique convergence to a fixed point and the rest of the parameter space.  We then compare the analytic and numerical simulations and demonstrate that the analytic calculation seems to be reasonably accurate, given the magnitude of the finite-size effects.  Finally we use the analytic calculation to assess the behavior in the limit where $N \to \infty$ and $p$ is large.

\section{Numerical exploration of the parameter space}\label{sec:behavior}

\subsection{Overview}

In this section we give an overview of our numerical exploration of the parameter space.   As observed by Galla and Farmer,  the strategies of the players can converge to any of the possible types of attractors, including fixed points, limit cycles and chaos.  In some regimes a given game may have multiple fixed points, i.e. multiple basins of attraction, but we have not observed this when the attractors are limit cycles or chaos.  In Fig.~\figr{lc_chaos_time_series} we show some examples.  The chaotic behavior can be low dimensional, as shown in the middle row, or high dimensional, as shown in the last row.  For high-dimensional chaos a given action typically has epochs in which it is almost never selected and others in which it is used frequently -- the range of variation is striking.  

When the system converges to a fixed point, it usually does so rather quickly, as shown in Fig.~\figr{fixed_point_time_series}(a).  However there are sometimes long metastable chaotic-looking transients that suddenly collapse to a fixed point, as shown in Fig.~\figr{fixed_point_time_series}(b).  This is particularly likely for small values of $\alpha/\beta$ and small positive $\Gamma$ (i.e., for weakly positively correlated payoffs and players with long memories).

\begin{figure}
\includegraphics[scale=0.5]{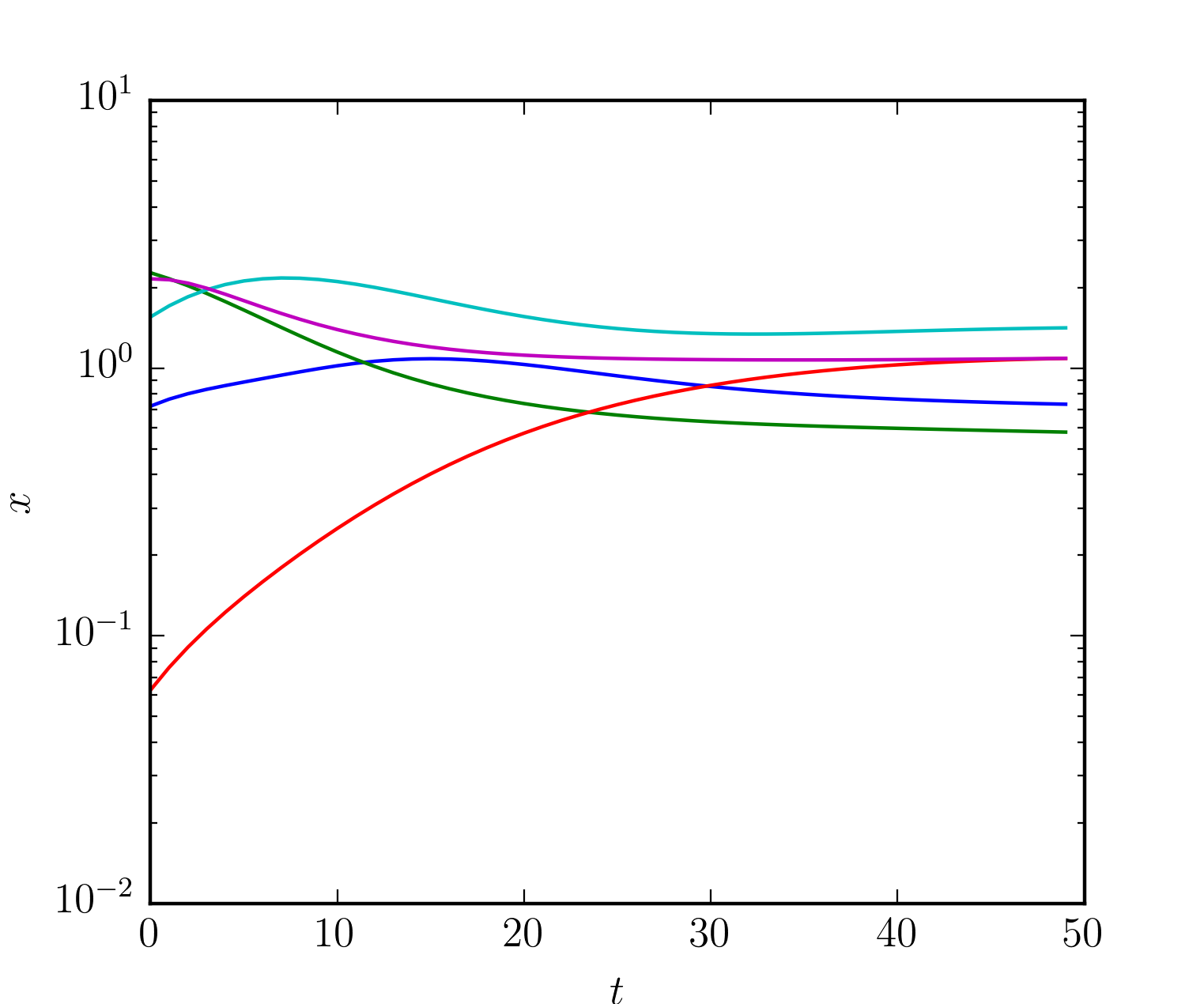}
\includegraphics[scale=0.5]{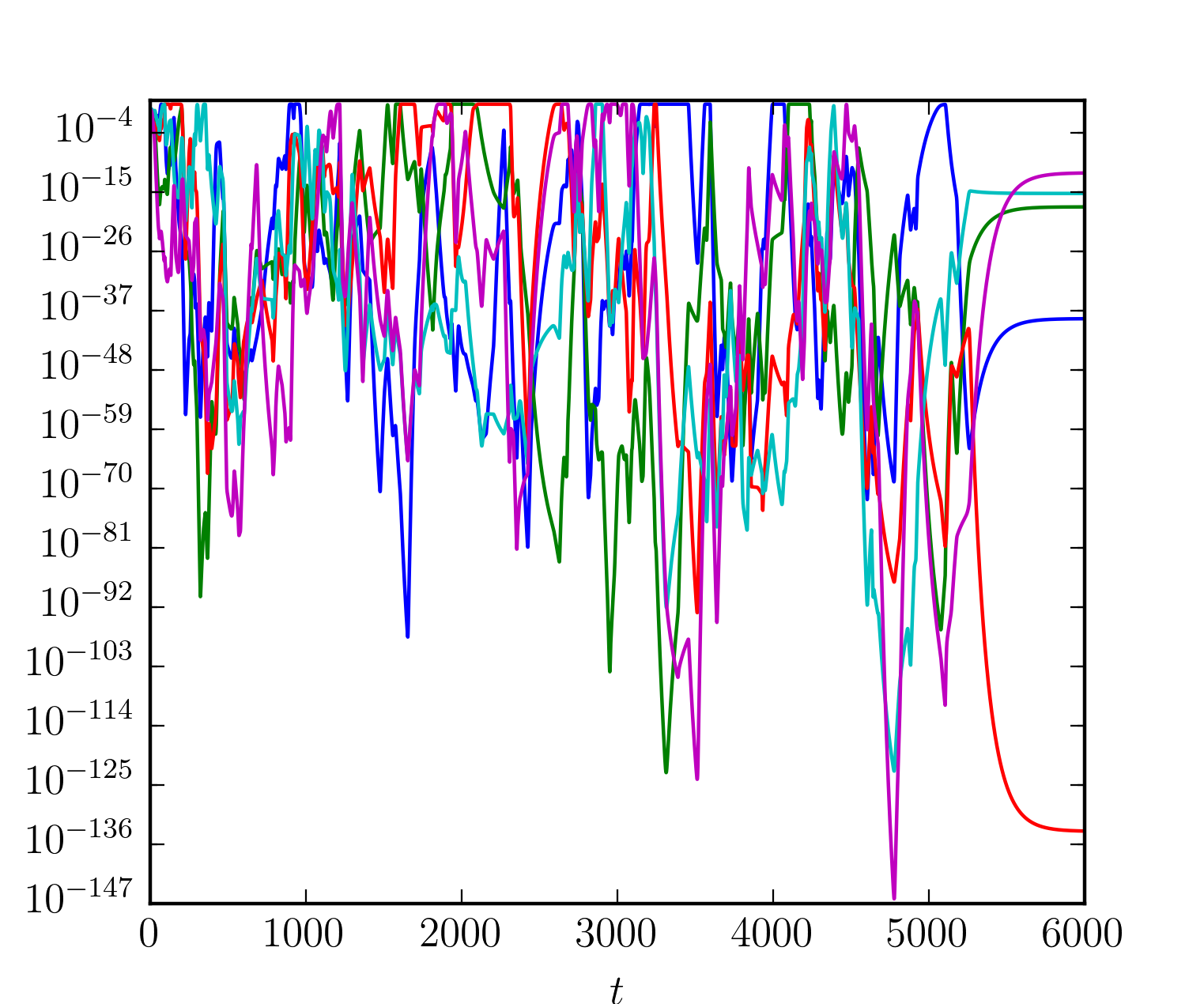}
\caption{Trajectories for EWA system leading to a fixed point in a three-player game. Panel (a) shows an instance in which a fixed point is reached relatively quickly. Panel (b) illustrates a metastable chaotic transient eventually collapsing to a fixed point. In both examples each player has a choice of $N = 20$ possible actions and the intensity of choice is $\beta = 0.05$. A random sample of five of the players' strategy components $x^\mu_i$ are plotted.  Remaining parameters are $\alpha = 0.1$, $\Gamma = -0.5$)  in panel (a), and $\alpha = 0.01$, $\Gamma = 0.1$ in panel (b).   }
  \figl{fixed_point_time_series}
\end{figure}

\begin{figure}
  \centering
\includegraphics[scale=0.7]{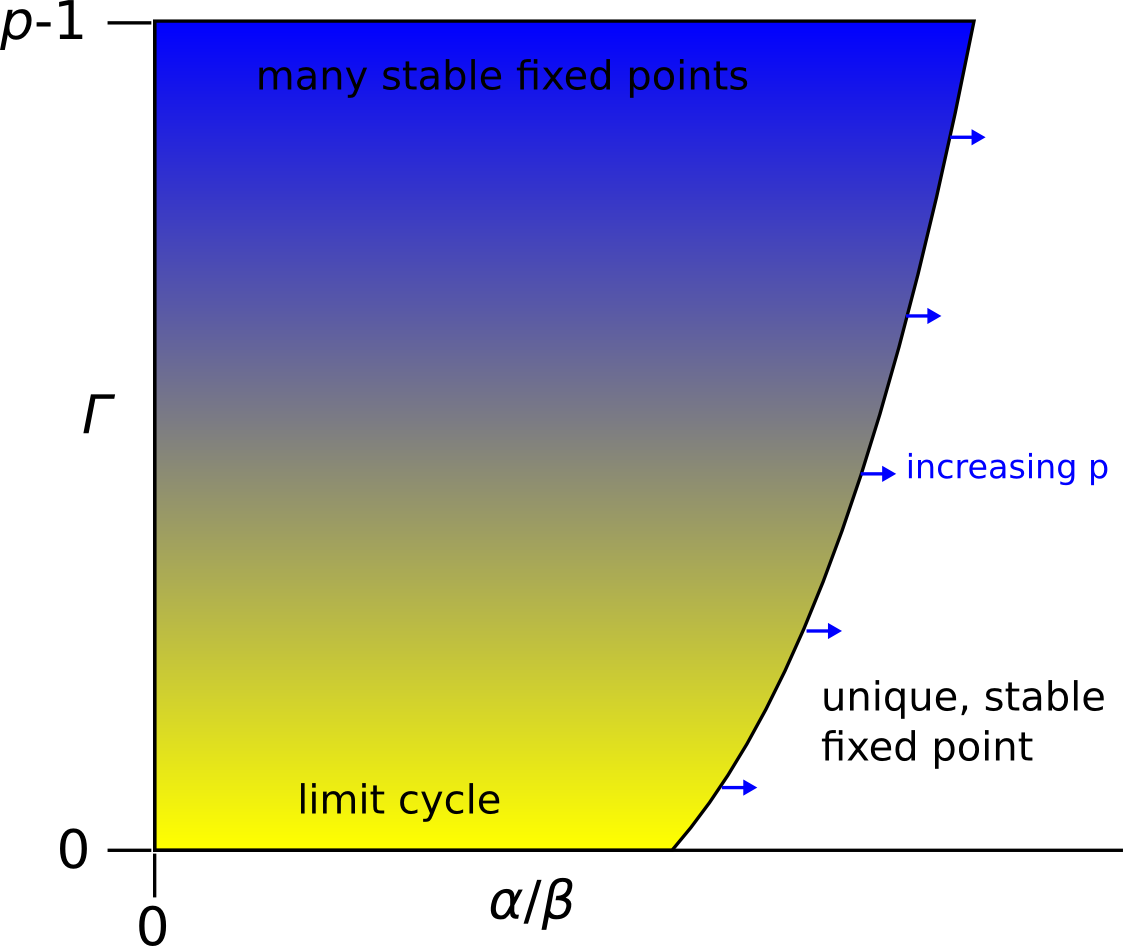}
\\
 (a) Positive $\Gamma$
\vspace{2em}

\includegraphics[scale=0.7]{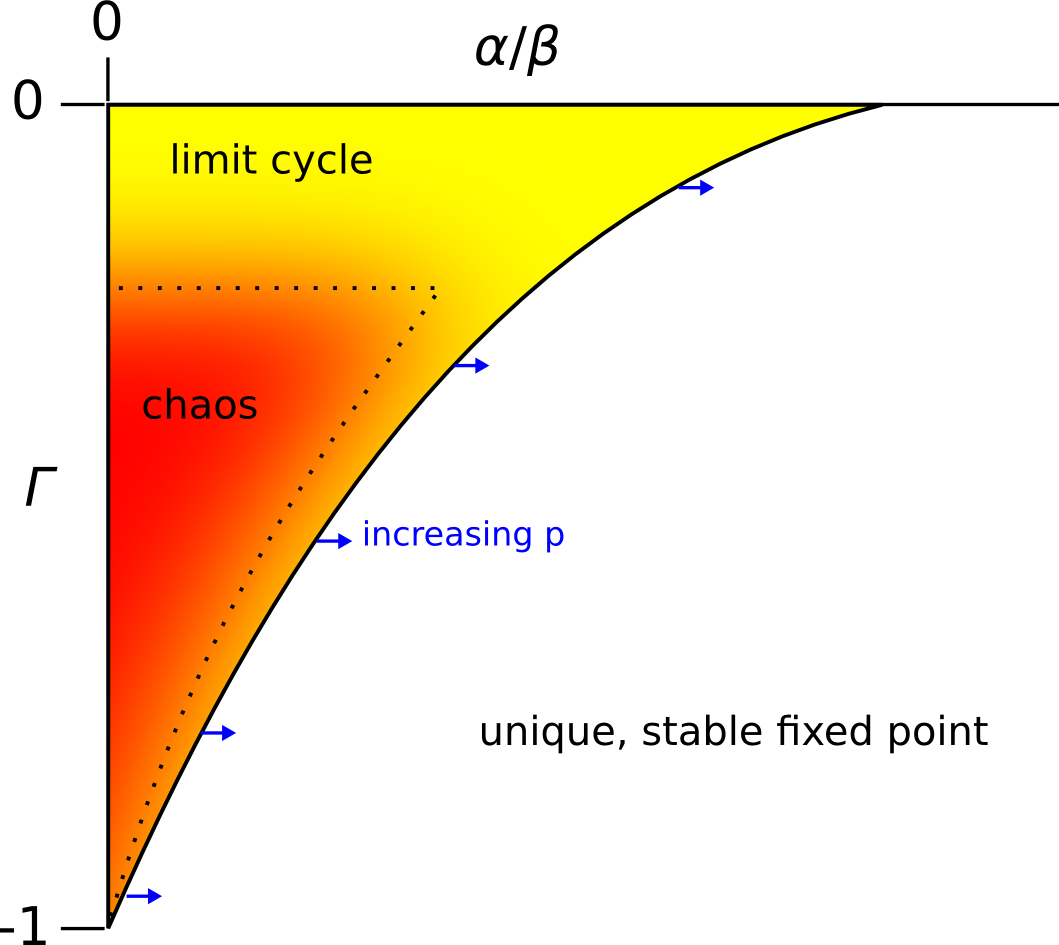}
\\
(b) Negative $\Gamma$
\\
  \caption{Schematic phase diagrams describing the observed long-term behavior of the $p$-player EWA system for large but finite~$N$.  In (a)  $\Gamma > 0$, meaning players' payoffs are positively correlated.  Here we observe a unique stable equilibrium for large $\alpha/\beta$ and multiple stable equilibria for small $\alpha/\beta$.   In (b) $\Gamma<0$, meaning players' payoffs are anti-correlated.  Here we once again observe a unique stable equilibrium for large $\alpha/\beta$, but we now observe chaos for small $\alpha/\beta$.  Limit cycles are common near the boundaries, particularly near $\Gamma \approx 0$.  We show the case for $p = 2$; as the number of players is increased the stability boundaries move to the right, but little else changes.  The heat map is subjective.  \figl{phase_diagrams}}
  
\end{figure}

Although the boundaries can be a bit fuzzy, the parameter space divides into distinct regions.  These are illustrated in the schematic diagram in Fig. \figr{phase_diagrams}.  We briefly describe the main features:
\smallskip

{\em Positively correlated payoffs:} For $\Gamma > 0$ and large $\alpha/\beta$ the dynamics converge to a unique fixed point. Holding $\Gamma$ constant, for small $\alpha/\beta$ the dynamics converges to one of multiple fixed points, see also Fig. \figr{pos_multiplicity_heat_maps}.
\smallskip

{\em Negatively correlated payoffs:} For $\Gamma < 0$ and large $\alpha/\beta$ the dynamics converge to a unique fixed point (just as for $\Gamma > 0$). Holding $\Gamma$ constant, for small $\alpha/\beta$ the dynamics are chaotic.   This corresponds to longer memory.  As $\Gamma$ increases from $\Gamma = -1$ to $\Gamma = 0$ the size of the chaotic region increases. The stability boundary dividing the region with complex dynamics from the unique equilibrium shifts to the right (i.e. toward a larger value of $\alpha/\beta$) as the number of players increases.  
\smallskip

{\em Uncorrelated payoffs:} Limit cycles are common near the boundaries, particularly near the boundary where $\Gamma \approx 0$, see also Figs. \figr{lc_heat_maps} and \figr{pos_lc_heat_maps} in the Appendix. In general the behaviors reported here are not strict, in the sense that generating different payoff matrices with the same values of $\alpha/\beta$ can result in different behaviors, particularly near the boundaries.  We conjecture that these are finite size effects, so that the boundaries would become distinct and the behavior at a given set of parameters would become crisp in the limit as $N \to \infty$. 
\smallskip 
 
{\em Prevalence of chaotic dynamics:} The key result is that as the number of players increases the size of the region with complex dynamics grows.  If $\Gamma > 0$, this means that the region with multiple fixed points becomes larger; if $\Gamma < 0$ this means that the region with chaotic dynamics becomes larger.  More players make the system less likely to converge to a unique equilibrium.  This is particularly important for zero sum games ($\Gamma = -1$); in this case for $p = 2$ chaos is only observed in the limit as $\alpha/\beta \to 0$,  whereas for $p > 2$ chaos is observed over a finite interval in $\alpha/\beta$.
\\

As already mentioned, it is not possible to perform numerical experiments for large values of both $p$ and $N$.  We will present some numerical evidence for these results, but they make more sense when guided by the theory.

\section{Generating functional approach}\label{sec:gfa}
We now turn to treat the learning dynamics analytically. In the language of the theory of disordered system the random payoff matrix in our problem represents quenched disorder. Techniques from spin glass physics can be applied to study the thermodynamic limit (i.e, the limit of large payoff matrices, $N\to\infty$). We use the Martin-Siggia-Rose generating functional to derive an effective dynamics. For a recent review of these methods see \cite{sollich}.

Broadly speaking the calculation proceeds as follows: in a first step the average over the disorder (the random payoff matrices) is carried out and the effective dynamics is derived. This process is subject to coloured noise, and reflects the statistics over all games within the Gaussian ensemble. In a second step, a fixed point of this dynamics is assumed.  We investigate the linear stability of this fixed point, and calculate the boundary to the phase in parameter space where more complex dynamics are seen.  Thus we cannot calculate where the dynamics follow limit cycles or chaos, but we can calculate the boundary between the unique stable fixed point and other behaviors.  As discussed later, we can only do this for $\Gamma < 0$.  We note that similar calculations have been carried out for replicator dynamics on two-player and in multi-player games \cite{opper, gallajpa}, see also \cite{oliveira} for approaches to $p$-player random games using static replica methods. 
 
 \subsection{Effective process}
The first step is to set up a generating functional to describe the probability measure of all possible paths of the dynamics. Performing the average over the assignment of payoff matrices an `effective dynamics' can then be derived. This calculation is based on path integrals, and somewhat lengthy. We do not report it here in the main paper, instead details are relegated to Appendix~\secr{pathint}. The outcome of the calculation is a stochastic integro-differential equation for the evolution of the distribution of components $x(t)$ of the players' strategies in the large-$N$ limit. This process is of the form
\BE
  \frac{\dot{x}(t)}{x(t)} &=& \Gamma \inti{t'} G(t,t') C(t,t')^{p-2} x(t') \nonumber \\
  &&- \frac{1}{r} \ln x(t) - \rho(t) + \eta(t),\label{eq:eff}
\EE
where $\eta(t)$ is a colored Gaussian random variable satisfying $\mean{\eta(t) \eta(t')}_* = C(t,t')^{p-1}$ and $\mean{\eta(t)}_* = 1$.  We use $\mean{\cdot}_*$ to denote an average over realizations of the effective dynamics. The dynamical order parameters $C(t,t')$ and $G(t,t')$ are correlation and response functions of the learning dynamics. They are determined  from
\be
 G(t,t') = \mean{\adiff{x(t)}{\eta(t')}}_*, ~~ C(t,t') = \mean{x(t)x(t')}_*.\label{eq:cg}
\ee
The effective process in Eq. (\ref{eq:eff}) together with Eqs. (\ref{eq:cg}) define a self-consistent system for $C$ and $G$. The function $\rho(t)$ in the effective process is a Lagrange multiplier ensuring normalization. It is defined such that $\mean{x}_* = 1$.  Note that in the derivation of the effective dynamics, it is assumed that each component of each player's strategy is initially drawn from an identical distribution.

\subsection{Fixed point solution}
We now focus on the dynamics at large values of $\alpha/\beta$. Numerical simulations of the learning dynamics suggest that one unique stable fixed point is found for any one realization of the game in this regime. We therefore make a fixed point ansatz for the effective dynamics. In such a stationary fixed point regime the response function $G(t,t')$ becomes a function of the time difference only, i.e., $G(t,t') = G(t-t')$, while the correlation function tends to a constant, $C(t,t') \equiv q$, see also \cite{opper,gallajpa,GallaFarmer} for further details.  Within the fixed-point ansatz the random variable $\eta(t)$ in Eq. (\ref{eq:eff})  tends to a constant value drawn from a Gaussian distribution with zero mean and variance $q^{p-1}$. Fixed points of the effective dynamics are then found from
\be
  x^* \plr{\Gamma q^{p-2} \chi x^* - \frac{1}{r} \ln x^* + \eta^* - \rho^*} = 0,
\ee
where $q = \mean{(x^*)^2}_*$ and $\chi = \intlu{\tau}{0}{\infty} G(\tau)$, and $x^*$, $\eta^*$, and $\rho^*$ are the fixed point values of $x$, $\eta$, and $\rho$, respectively.  We can write $\eta^* = q^{(p-1)/2} z$, where $z$ is a standard Gaussian random variable.  Then, dropping the stars, we have
\begin{equation}\label{eq:fixedpointeqn}
  x(z) \plr{\Gamma q^{p-2} \chi x(z) - \frac{1}{r} \ln x(z) + q^{\frac{p-1}{2}} z - \rho} = 0,
\end{equation}
where $\chi$, $q$, and $\rho$ are to be determined from
\BE
  \mean{\pdiff{x(z)}{z}}_* &=& q^{\frac{p-1}{2}} \chi , \nonumber \\
  \mean{x(z)^2}_* &=& q, \nonumber \\
  \mean{x(z)}_* &=& 1.
\EE
These relations can be re-written as
\BE
  \intDlu{z}{-\infty}{\infty} \pdiff{x(z)}{z} &= &q^{\frac{p-1}{2}} \chi , \nonumber \\
  \intDlu{z}{-\infty}{\infty} x(z)^2 &=& q, \nonumber \\
  \intDlu{z}{-\infty}{\infty} x(z) &=& 1,\label{eq:integral}
\EE
where $\DD z = \frac{\dd z}{\sqrt{2 \pi}} \e^{-z^2/2}$.

The relation in Eq. (\ref{eq:fixedpointeqn}) can be re-arranged to give an explicit expression for $x(z)$ in terms of the so-called Lambert W function $W(\cdot)$. The value $W(y)$ is defined as the solution of the equation $We^W=y$. Restricting $W$ and $y$ to the real line, the solution exists for $y\geq -1/e$. It is uniquely defined for $y\geq 0$ and double valued for $-1/e<y<0$. We find
\be  \label{eq:xfp}
x = - \frac{1}{\Gamma r q^{p-2} \chi} W\left(-\Gamma r q^{p-2} \chi e^{r(q^{(p-1)/2}z - \rho)}\right).
\ee

We note that it is not clear that Eq. (\ref{eq:xfp}) has valid solutions for all choices of the model parameters. If these do not exist the fixed point ansatz is invalid, and so we do not expect the dynamics to settle down. There may also be instances in which Eq. \ref{eq:fixedpointeqn} has multiple solutions for $x$ for a given value of the standard Gaussian variable $z$. In principle the distribution of fixed points could be composed of any mixture of these solutions. If the argument of the Lambert function is positive however, there is a unique and well defined solution, $x(z)$, for any value of $z$. Throughout this discussion it is important to keep in mind that the macroscopic order parameters $q,\chi$ and $\rho$ are to be determined self-consistently via Eqs. (\ref{eq:integral}).

\subsection{Stability analysis} \secl{stabcurves}
By numerically solving the fixed point equations, we see that for a given value of $p$, stable fixed points exist for large values of $1/r$ but not for small values. We now proceed to determine the boundary of stability. Suppose the effective process in Eq. (\ref{eq:eff}) is perturbed from a fixed point by a small noise term~$\xi(t)$. We then have
\BE
  \dot{x}(t) &= &x(t) \bigg[\Gamma \inti{t'} G(t,t') C(t,t')^{p-2} x(t') \nonumber \\
&&  - \frac{1}{r} \ln x(t) - \rho(t) + \eta(t) + \xi(t)\bigg].
\EE
We assume that $\xi(t)$ is white Gaussian noise of unit amplitude. Writing $x(t) = x^* + \htx(t)$ and $\eta(t) = \eta^* + \hteta(t)$, and keeping only linear terms in $\xi$, $\htx$, and $\hteta$, we obtain
\BE \eql{linearisednoise}
  \dot{\htx}(t) &= &- \frac{1}{r} \htx(t) + x^* \bigg[\Gamma \inti{t'} G(t-t') C(t-t')^{p-2} \htx(t')\nonumber \\
  && + \hteta(t) + \xi(t)\bigg].
\EE

Defining $H(t,t') = G(t,t') C(t,t')^{p-2}$, and taking the Fourier transform of Eq. \eqr{linearisednoise}, yields
\begin{equation}
  \plr{ \frac{\i \omega + r^{-1}}{x^*} - \Gamma \tiH(\omega)} \tix(\omega) = \tieta(\omega) + \tixi(\omega),
\end{equation}
where the tildes denote Fourier transforms.  This leads to the relation
\BE
  \mean{\mlr{\tix(\omega)}^2}_* &=& \plr{(p-1)\mean{(x^*)^2}_*^{p-2} \mean{\mlr{\tix(\omega)}^2}_*+ \mean{\mlr{\tixi(\omega)}^2}_*}\nonumber \\
  &&\times \mean{\mlr{A(\omega,x^*)}^{-2}}_*,
\EE
where 
\begin{equation}
  {\cal A}(\omega,x^*) = \frac{\i \omega + r^{-1}}{x^*} - \Gamma \tiH(\omega).
\end{equation}
We can write this as
\begin{equation}
  \mean{\mlr{\tix(\omega)}^2}_* = \plr{\mean{\mlr{{\cal A}(\omega, x^*)}^{-2}}_*^{-1} - (p-1) q^{p-2}}^{-1}.
\end{equation}
The left-hand side is positive by definition so the calculation runs into a contradiction if the expression on the right-hand side turns negative. As it approaches zero (from above) the magnitude of fluctuations diverges. The fixed point can only be stable when
\begin{equation}
  \mean{\mlr{{\cal A}(\omega,x^*)}^{-2}}_* < \frac{1}{(p-1) \mean{(x^*)^2}_*^{p-2}}.
\end{equation}
Following \cite{opper} we focus on $\omega=0$, so stability can only be expected provided that
\begin{equation}
  \mean{\mlr{\frac{1}{r x^*} - \Gamma q^{p-2} \chi}^{-2}}_* < \frac{1}{(p-1) q^{p-2}},
\end{equation}
i.e.,
\begin{equation} \label{eq:finalstability}
  \intDlu{z}{-\infty}{\infty} \plr{\frac{1}{r x(z)} - \Gamma q^{p-2} \chi}^{-2} \leq \frac{1}{(p-1) q^{p-2}}.
\end{equation}

For negative values of $\Gamma$, the position of the stability boundary can be determined straightforwardly by numerically solving Eqs. (\ref{eq:integral},\ref{eq:xfp}) for a given set of model parameters, and by subsequently evaluating the stability condition Eq. (\ref{eq:finalstability}) throughout parameter space.  As already mentioned, this procedure fails for $\Gamma > 0$.

\subsection{The large-\texorpdfstring{$p$}{p} limit}
In general the location of the stability region defined by (\ref{eq:finalstability}) cannot be determined fully analytically.  However, it is possible to make progress in several limits as shown below and make a good sketch of the behavior when $\Gamma < 0$.

Taking $\Gamma\to 0$ in Eq. (\ref{eq:fixedpointeqn}) we find $x(z)=\exp\left[r(q^{(p-1)/2}z-\rho)\right]$, and using Eq. (\ref{eq:integral}) the order parameters $r$, $q$, and $\rho$ satisfy
\BE
  \chi &= &r, \nonumber \\
  \exp(r^2 q^{p-1}) &=& q,\nonumber \\
  \rho &=& \frac{\ln q}{2 r},
\EE
in this limit.
The second expression is equivalent to
\begin{equation}
  q = \plr{- \frac{W\plr{-(p-1)r ^ 2}}{(p-1)r ^ 2}}^{1/(p-1)},
\end{equation}
where $W(\cdot)$ is the Lambert W function.  For $(p-1) r^2 > 1/e$ this has no solutions.  For $(p-1)r^2 < 1/e$ it has two, given by the two branches of the Lambert W function.  The solution corresponding to the upper branch ($W > -1$) satisfies (\ref{eq:finalstability}) so is always stable, while the other is always unstable.  Therefore, on the~$\Gamma = 0$ line, there is a stable fixed point only for large values of $1/r$, with the stability boundary given by
\BE
  r = \chi &= &\frac{1}{\sqrt{(p-1) \e}}, \nonumber \\
  q &= &\exp\plr{\frac{1}{p-1}}, \nonumber \\
  \rho &= &\frac{1}{2} \sqrt{\frac{\e}{p-1}}. \label{eq:stabgamma0}
\EE

For $\Gamma=-1$ and $p=2$ the system is stable as soon as there is non-zero memory loss ($\alpha>0$), that is to say the stability line passes through the point $(1/r=0, \Gamma = -1)$ when $p = 2$, see also \cite{GallaFarmer}.  For larger numbers of players we find that the stability boundary never reaches the $1/r=0$ line.

 The boundary between the two regions crosses the $\Gamma = 0$ line at the location we have determined analytically, and tends to a straight line as $p \rightarrow \infty$, as shown in Fig. \ref{fig:stabcurves}.  It is in fact possible to demonstrate this analytically, see Appendix~\secr{straightlines} for details.

\begin{figure}
  \centering
\includegraphics[scale=0.6]{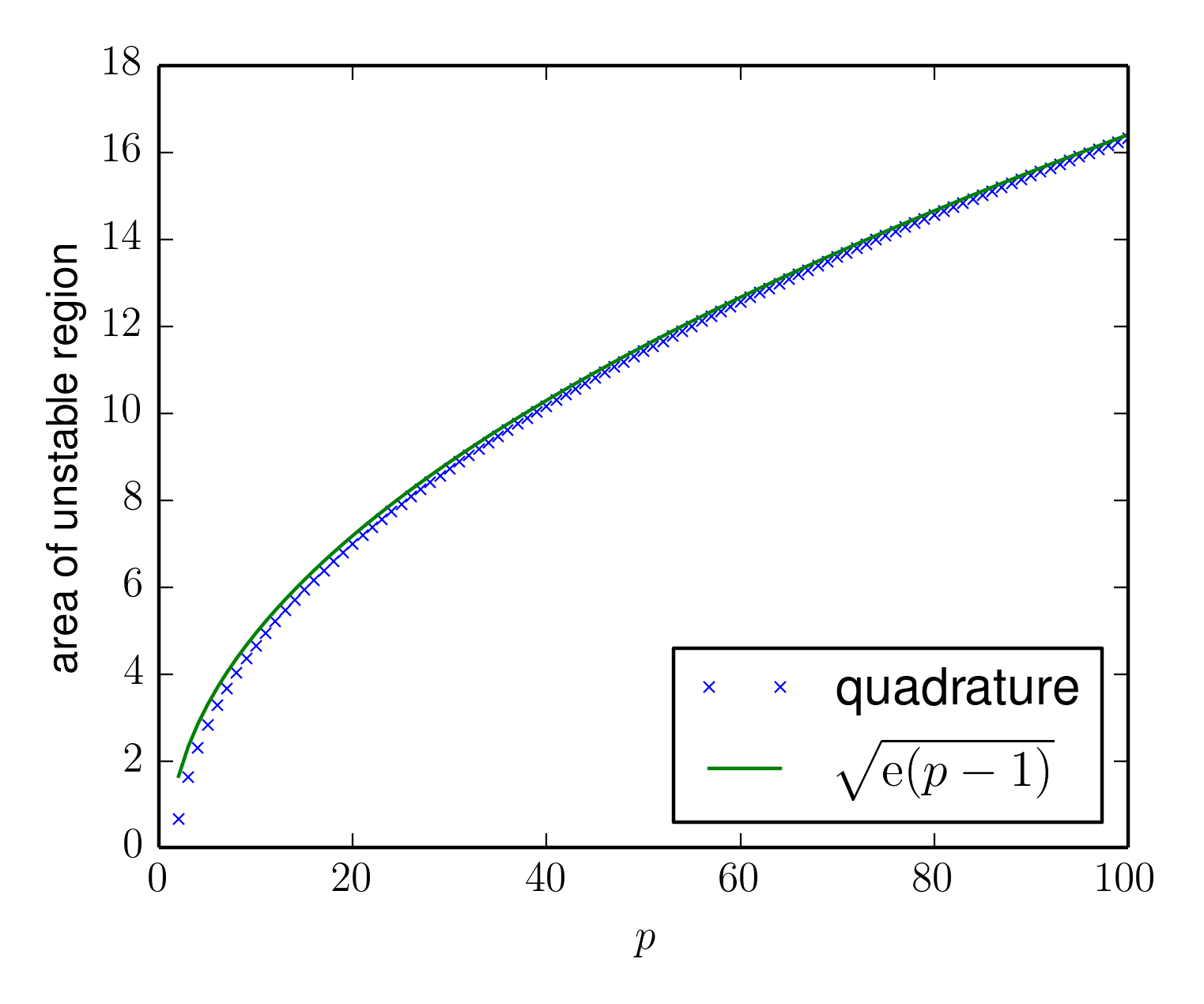}
\caption{Plot showing the area of the unstable region for negative~$\Gamma$ as a function of the number of players, $p$.  This area is estimated numerically using Gaussian quadrature on results obtained for $\beta=0.01$; this is valid in the limit as $N \to \infty$. The analytic estimate of the area is $\sqrt{\e (p-1)}$, see Eq. (\ref{eq:size}). This indicates that the area of the parameter space with complex dynamics goes to infinity proportional to $\sqrt{p}$ as $p \to \infty$.  \label{fig:stabcurvesarea}}
\end{figure}

Based on this result we can  estimate that the size, $A$, of the unstable region in the $\alpha/\beta-\Gamma$ plane as shown in Fig. \ref{fig:stabcurvesarea} (restricted to $-1\leq\Gamma\leq 0$). As the number of players $p$ increases this area grows as
\be\label{eq:size}
A\approx  \sqrt{\e (p-1)}
\ee
as shown in Fig. \ref{fig:stabcurvesarea}.

Thus, in the case where $\Gamma < 0$ and $p \to \infty$, the unstable region takes over the entire parameter space.  That is, in situations where the players' payoffs are anticorrelated, their behavior will always be complex and will never settle down on an equilibrium.

\begin{figure}
  \centering
  \includegraphics[scale=0.6]{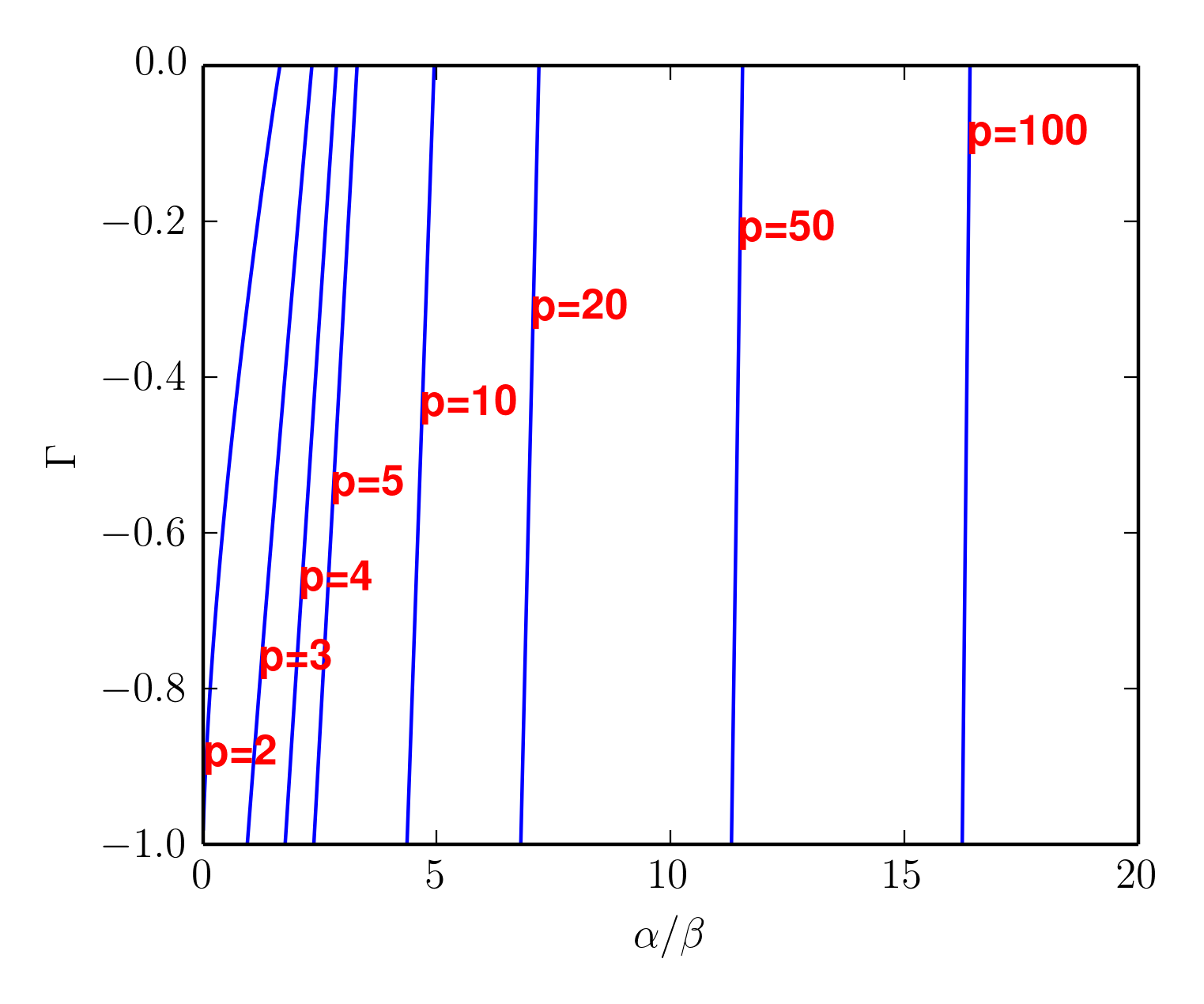}
  \caption{Stability boundaries of the effective dynamics for several values of $p$ as a function of $\alpha/\beta$ and $\Gamma$, for the case where $\Gamma<0$.  Each curve is the stability boundary for the stated value of $p$.  To the left of any curve the fixed point of the effective dynamics is unstable, to the right it is stable.  The key result is that the stability boundary moves to the left as $p$ increases, so the size of the regime with complex dynamics grows.  \label{fig:stabcurves}}
\end{figure}

\section{Comparison between theory and numerical experiments}

We now compare the numerical results to the theoretical predictions.  We measure the stability boundary in the numerical experiments by determining whether or not the system converges to a unique fixed point, independent of initial conditions.  To do this we choose a set of parameters and initial conditions, iterate the dynamics for a large number of time steps, and apply heuristics to check whether the players' strategies have converged to a fixed point.  If we find a fixed point, we repeat this for many different initial conditions and check to see whether   we always find the same fixed point.  We also perform a similar procedure to check for limit cycles.  The precise methods we use are discussed in detail in appendix~\secr{numerics}. 
\begin{figure}
\includegraphics[scale=0.45]{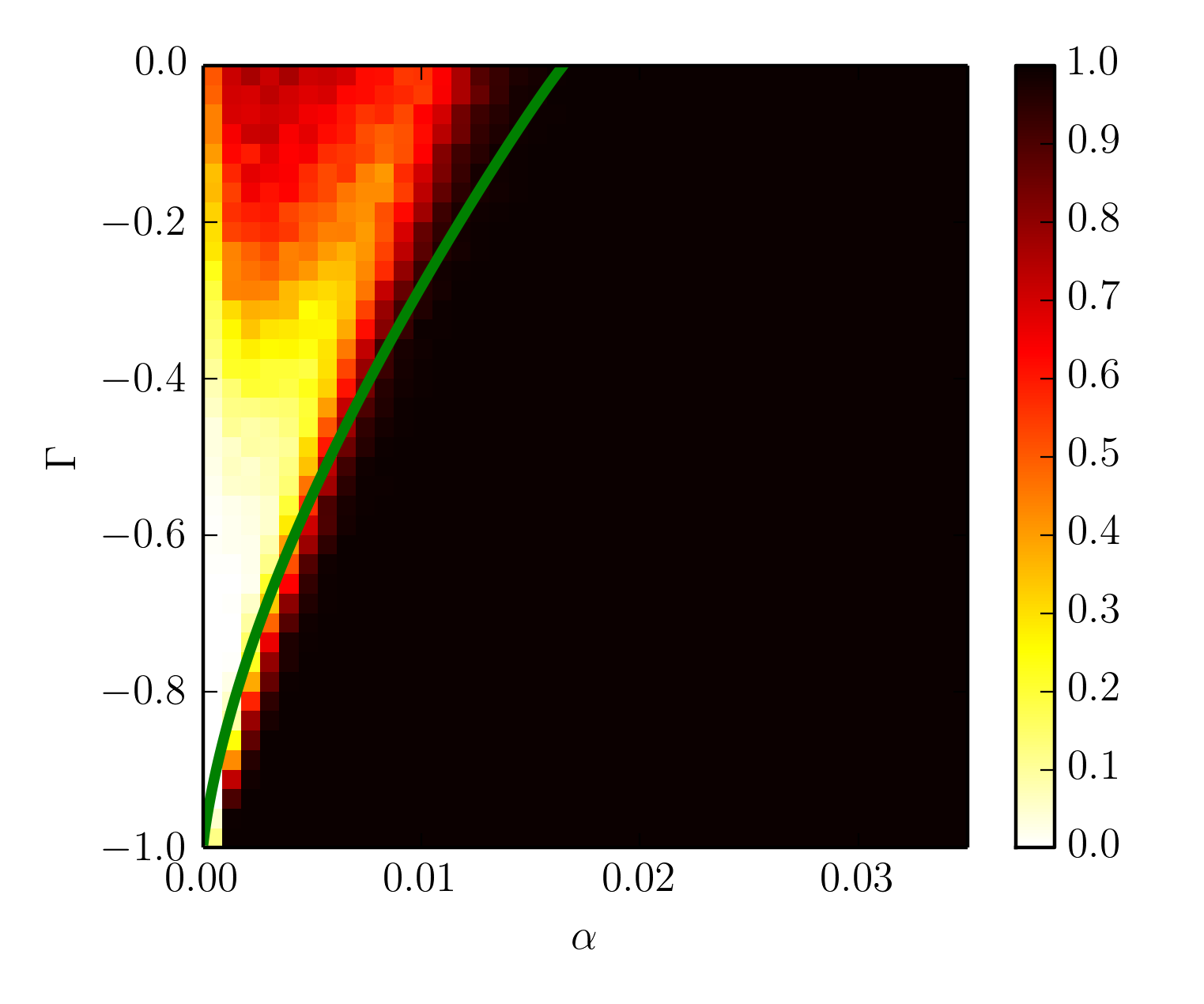}
\\
(a) $p=2$, $N=50$.
\\
\includegraphics[scale=0.45]{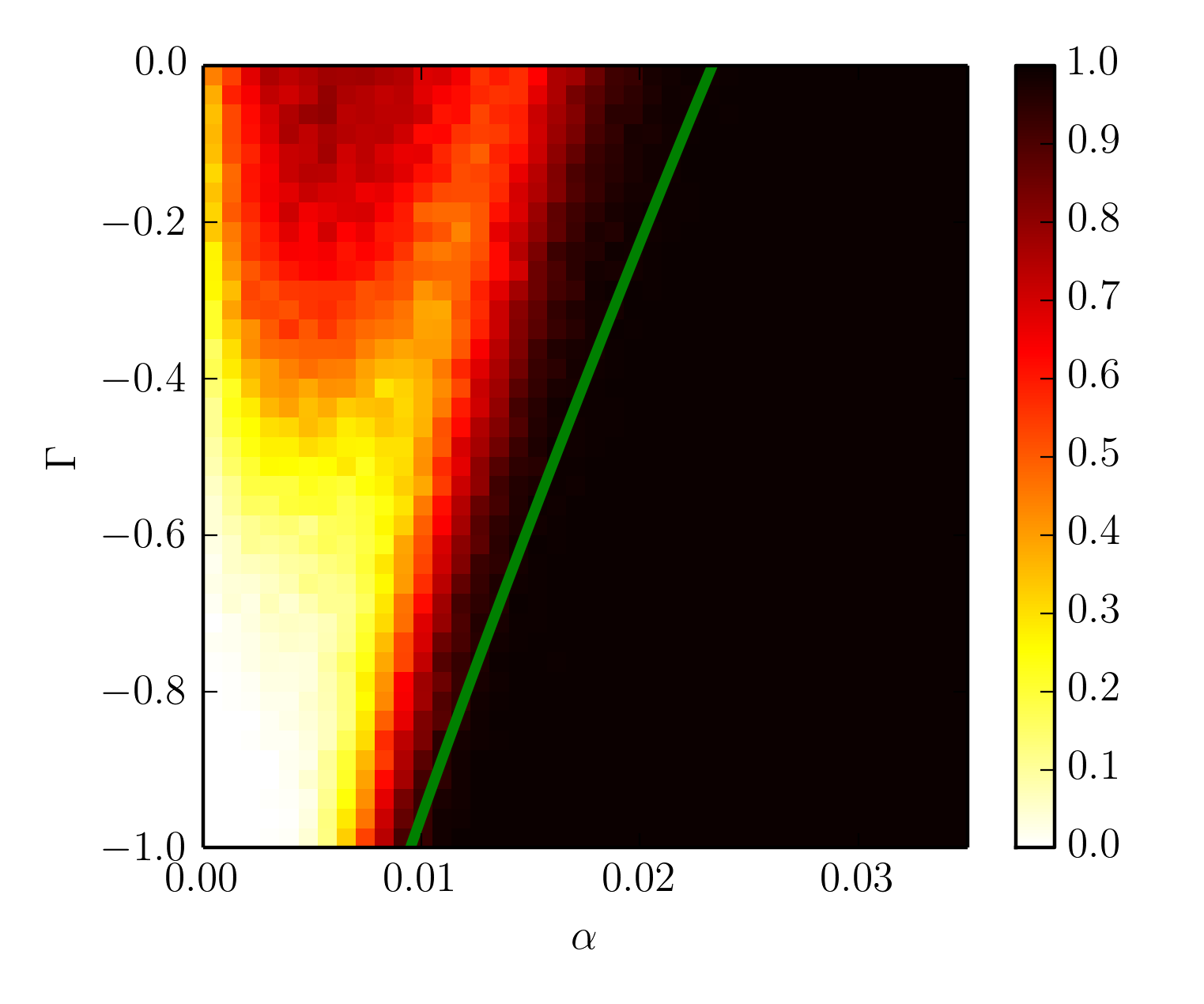}
\\
(b) $p=3$, $N=12$.
\\
\includegraphics[scale=0.45]{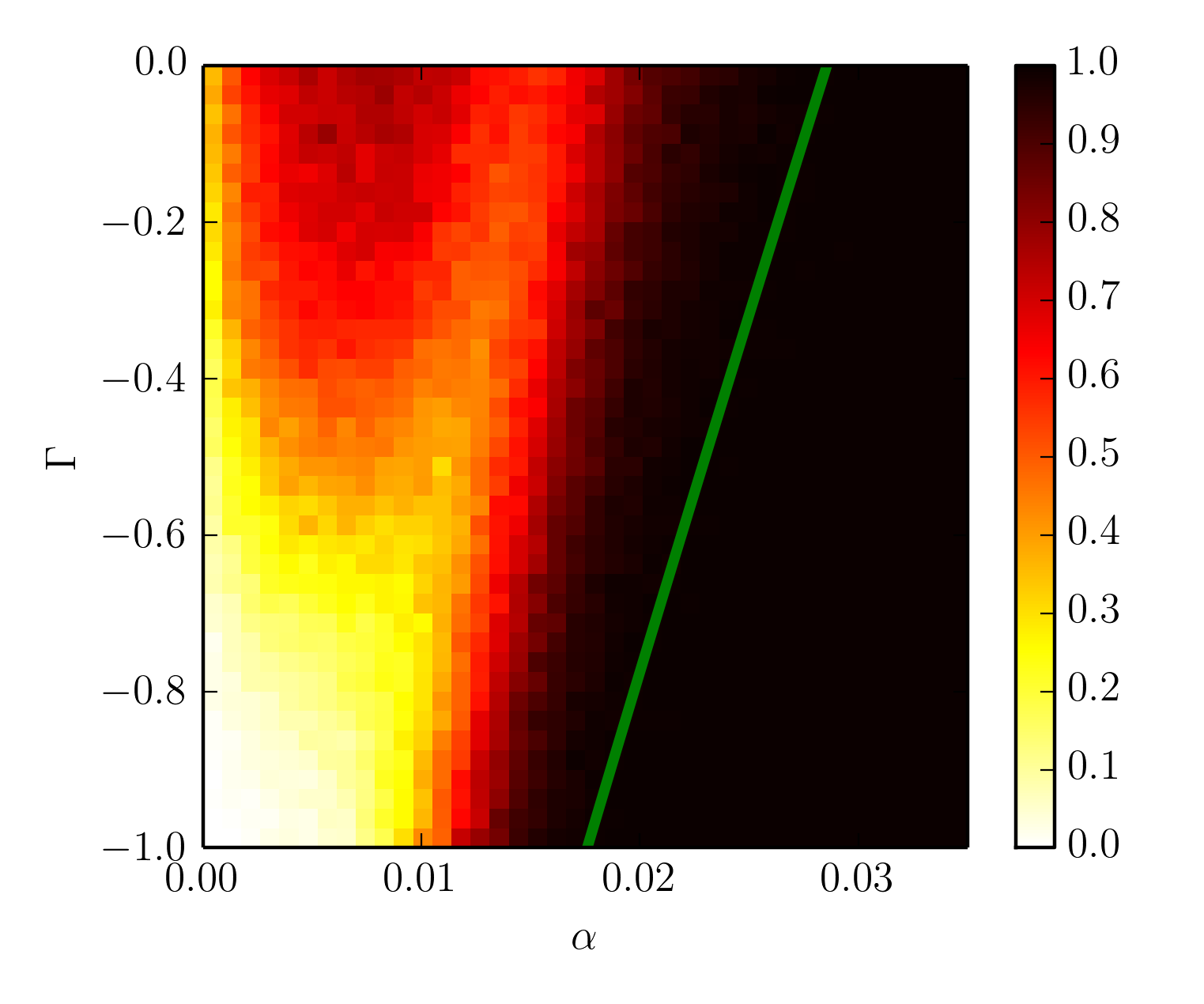}
\\
(c) $p=4$, $N=6$.
 \\
 \includegraphics[scale=0.45]{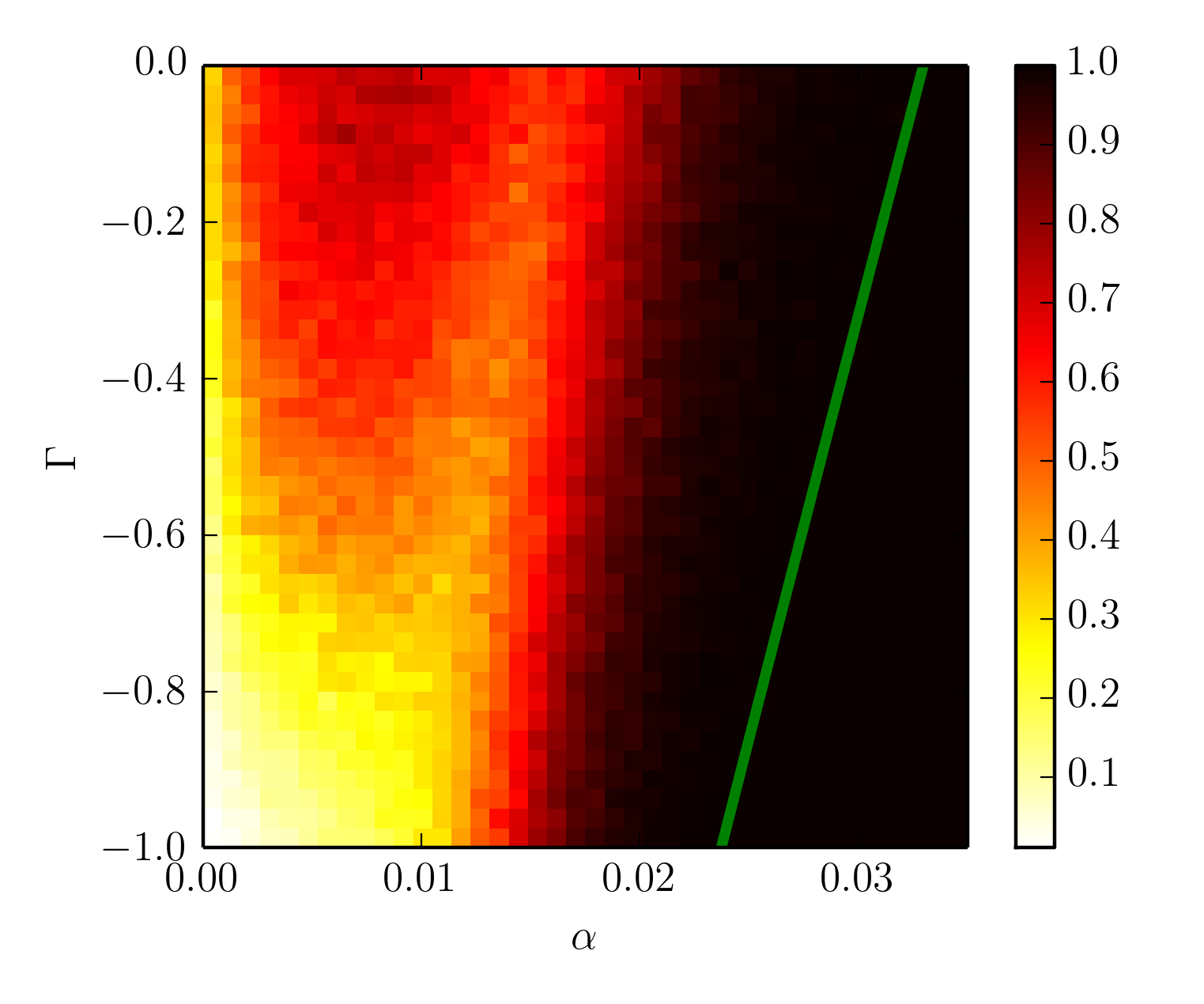}
 \\
(d) $p=5$, $N=4$.
\\
  \caption{Probability of convergence to a fixed point as a function of the memory parameter $\alpha$ and the competition parameter $\Gamma$. For each set of parameters we iterate the system from $500$ random initial conditions.  The heat maps show the fraction that converged to a  fixed point (convergence criteria discussed in Appendix \secr{numerics}).  Black means $100\%$ convergence, red (grey) indicates the majority converge, yellow a minority, and white no convergence.  The unstable region extends to larger values of $\alpha$ as the number of players is increased. The solid curves are derived from a generating functional analysis ($N\to\infty$, see  Sec.~\secr{stabcurves}), and separate the region in which a unique stable fixed point is to be expected in this limit (to the right of the green curves) from regions in parameters space where the behavior is more complex.}
  \figl{fp_heat_maps}
\end{figure}
 
Fig.~\figr{fp_heat_maps} shows how the likelihood of converging to a unique stable fixed point varies throughout the negative-$\Gamma$ region of parameter space for different values of $N$ and $p$.   The values we choose are roughly at the limit of what was computationally feasible.  We investigate $p = 2, 3, 4, 5$, which constrains the corresponding values of $N$ to be $N = 50, 12, 6, 4$.  We then sweep $\Gamma$ and $\alpha$ with $\beta = 0.05$ and construct a heat map showing the likelihood of convergence to a unique stable fixed point.  We then compare this to the stability line predicted by the generating functional approach described in the previous section.  

The heat map of Fig.~\figr{fp_heat_maps} is constructed so that black corresponds to convergence to a stable fixed point $100\%$ of the time, red (grey) to convergence roughly $50\% - 70\%$, yellow (light grey) $10\% - 35\%$, and white to the case in which unique stable fixed points are never found.   The behavior is consistent with what we described schematically in Fig.~\figr{phase_diagrams}(b):  Unique stable fixed points are more likely for higher $\alpha$ (i.e. short memory) and the size of the stable region grows with increasing $\Gamma$.   The region in which complex dynamics are observed grows as $p$ increases; in particular for the zero-sum case where $\Gamma = -1$ the size of the interval corresponding to complex dynamics is finite and growing with $p$.  

The correspondence between the predicted vs. the observed stability boundary gets better as $N$ increases.  For $p = 2$, where we can make $N = 50$, the correspondence is quite good (Fig.~\figr{fp_heat_maps}(a)); for $p = 5$, where we are only able to make $N = 4$, the correspondence is not as good (Fig. ~\figr{fp_heat_maps}(d)); the stability line scales more or less tracks the numerically-observed boundary, but is consistently to the right of it.  Given that $N = 4 \ll N = \infty$, it is not surprising that the approximation is not perfectly accurate.  We hypothesize that this is due to finite size effects.  To test this, in  Fig.~\figr{N_heat_maps} in the Appendix we hold the number of players constant at $p = 2$ and systematically vary $N$.  We find the correspondence between theory and experiment improving with increasing $N$.  In addition the behavior becomes crisper in the sense that the transition from certain convergence to a unique fixed point to never converging to a unique fixed point happens more suddenly when $N$ becomes large.   This indicates that the generating function methods gives reasonably good predictions for large $N$, lending confidence to its reliability in the limit as $N \to \infty$.

\section{Chaos and volatility clustering}\label{sec:vol}

Before concluding we would like to make a few notes about chaos and volatility clustering.  We have so far asserted that much of the behavior in the competitive region where $\Gamma < 0$ is chaotic, without presenting any evidence.  In fact we have done extensive computation of Lyapunov exponents using the procedures described in Galla and Farmer \cite{GallaFarmer}.  While we experience some numerical problems we can nonetheless state with confidence that the preponderance of the complex dynamics to the right of the stability line corresponds to chaos.  Problems arise because it can sometimes be difficult to numerically distinguish chaos and limit cycles without making very long simulations, and because of the metastable chaos observed in Fig.~\figr{fixed_point_time_series}, which means that in any given simulation there is a small but nonzero probability that the simulation will eventually collapse to a fixed point.  Nonetheless, most of the time we observe chaos, and as $p$ increases it tends to be of higher dimension.  To prove our main point here and compare to the theory we only needed to determine whether or not we observe convergence to a unique fixed point, which is much easier computationally, so we have chosen not to present evidence based on Lyapunov exponents.

\begin{figure}
  \centering
  \includegraphics[scale=0.6]{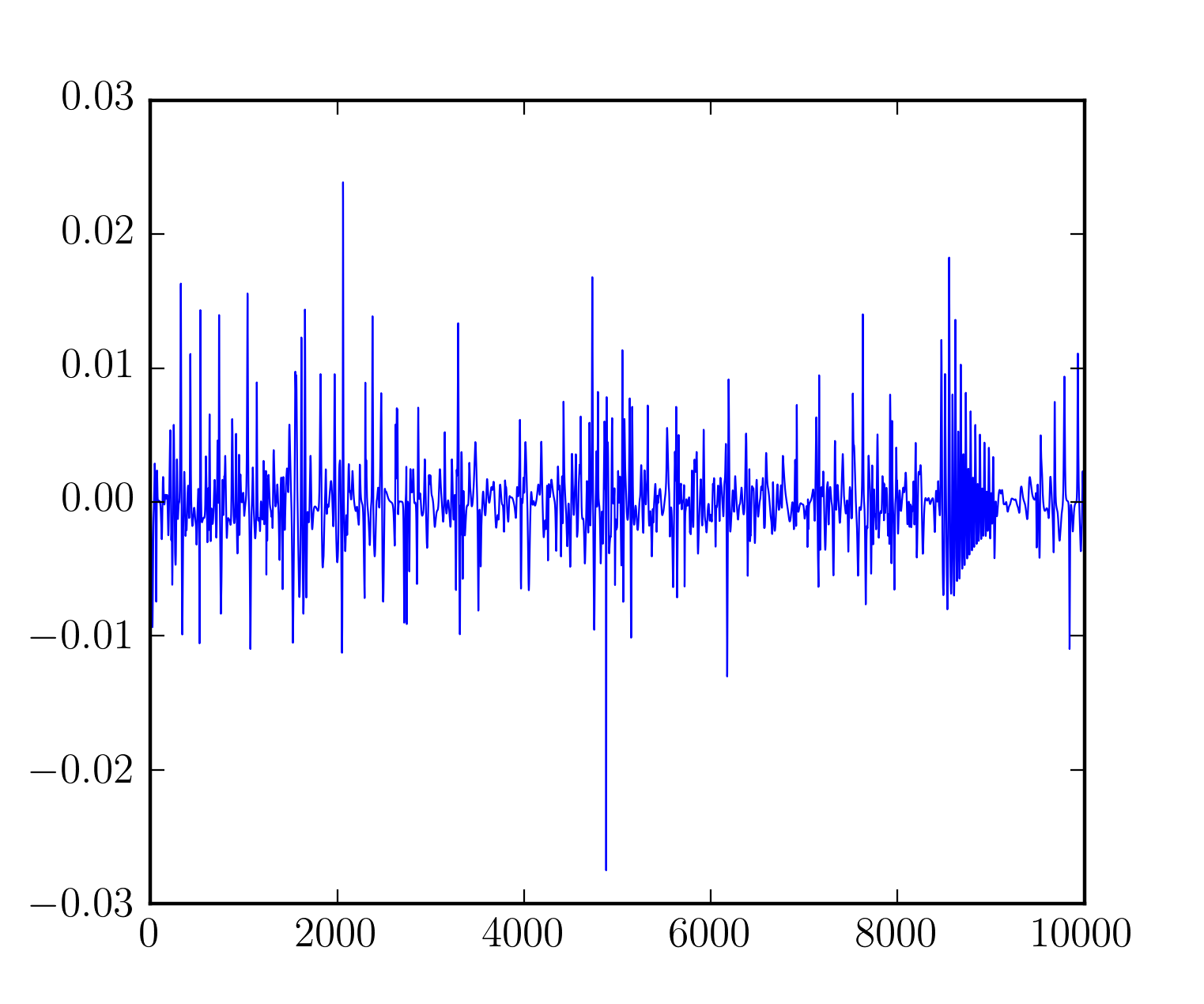}
  \caption{Time series of the changes in the sum of the players' payoffs for a game with three-players.  This corresponds to high dimensional chaos and clustered volatility.  By this we mean the tendency for time variability to be positively autocorrelated, with periods of relative calm and periods of relative variability.
  \label{fig:clustering}}
\end{figure}

In the chaotic regime we consistency observe clustered volatility, similar to that reported for $p = 2$ by Galla and Farmer.  By this we mean that the fluctuation in payoffs to the players fluctuates in time in a way that is ``clustered", i.e. positively autocorrelated.  There are epochs in which the payoffs are relatively steady and other epochs in which they are highly variable, as shown in Fig.~\ref{fig:clustering}. For $p > 2$ the chaos tends to be higher dimensional and the clustered volatility stronger.  Clustered volatility is common in many real-world situations, including financial time series; our work here suggests that this may be a generic result for games in which players learn their strategies using procedures similar to EWA.  We conjecture that this is connected to the tendency for a given action to vary from being used frequently for long periods of time to being almost never used, as observed in Fig.~\figr{lc_chaos_time_series}(c), but this remains to be investigated.

\section{Conclusions}
In summary we have  characterized the outcome of adaptive learning in complex multi-player games with Gaussian random payoff matrices. The learning dynamics we have simulated is a special case of experience-weighted attraction learning used in behavioral economics. We see different types of dynamical behavior: convergence to fixed points, limit cycles and chaotic trajectories. Broadly speaking, convergence to a unique stable equilibrium is observed when players' learning neglects outcomes from the distant past, i.e. when they forget quickly, corresponding to large values of $\alpha/\beta$.    In contrast, when they have a long memory, i.e. small $\alpha/\beta$, we observed more complex dynamics.  The nature of this dynamics depends on how competitive the game is.  For competitive games (i.e. with $\Gamma < 0$) the complex dynamics exhibits itself as a limit cycle or chaotic attractor. For cooperative games (i.e. with $\Gamma > 0$) the complex dynamics exhibits itself as a multiplicity of fixed points.  The boundaries between these behaviors become sharper as $N$ increases.

The main focus of this paper was to study the properties of games with more than two players.  For two players we replicate the results of Galla and Farmer \cite{GallaFarmer}.  For more than two players we are not able to simulate situations with a very large number of actions due to computational constraints, e.g. for $p = 3$ players the largest number of actions we simulated was $N = 12$.  To clearly understand the behavior for large $p$ and large $N$ we rely on an analytic treatment, which allows us to estimate the stability boundary in the limit as $N \to \infty$.  
Based on tools from the theory of disordered systems, we have carried out a generating-functional analysis of the continuous-time limit of EWA learning, and we have derived approximate semi-analytical results for the onset of stability for games with an infinite number of strategies and with an arbitrary finite number of players, $p$. These results reveal that the parameter range in which learning cannot be expected to settle down to fixed points increases as the number of players in the game grows. This is summarized in Figs. \ref{fig:stabcurves} and \ref{fig:stabcurvesarea}.  In contrast to the Galla and Farmer paper, where the analytic results were just making the numerical results more rigorous, here the analytic methods were essential to understand the behavior for large $p$.

In the introduction we posed our objective as seeking a ``Reynolds number" for estimating the {\it a priori} likelihood of complex dynamics, in much the same way that the Reynolds number characterizes turbulence in fluid flow.  Indeed the parameter $r = \beta/\alpha$ characterizing the timescale of the learning process does a reasonably good job in this context:  As $r$ gets bigger, complex dynamics becomes more likely.  The transition also depends on the competitiveness of the game, characterized by $\Gamma$, as well as the number of players.  Our key result here is that complex dynamics become more likely as the number of players increases.  This is not surprising given that games with more players are more complicated, and perhaps also more complex, in the sense that there are more factors to take into account and more inherent degrees of freedom. 

The standard theoretical approach in economics assumes convergence to an equilibrium from the outset.  Our results here suggest that under circumstances where the players have a long memory of the past, this approach may be inherently flawed.  This is particularly true when there are many agents.  Our results suggest that there may be large regimes in which the assumption of a unique equilibrium is completely invalid, and where approaches that can accommodate chaotic dynamics, such as agent-based modeling, are needed.  Of course in this paper we have only studied one family of learning algorithms, and we have focused on games with many actions.  More work is needed to give a definitive answer to the question above.

\section*{Acknowledgements}
JBTS thanks the Engineering and Physical Sciences Research Council (EPSRC) for support and JDF thanks the Institute for New Economic Thinking.

\clearpage

\appendix
\onecolumngrid

\renewcommand{\theequation}{A\arabic{equation}}
\renewcommand\thesection{A\arabic{section}}
\renewcommand\thefigure{A\arabic{figure}}
\setcounter{equation}{0}
\setcounter{section}{0}
\setcounter{figure}{0}
\newpage
\section*{Appendix}

\section{Generating functional analysis} \secl{pathint}
Following \cite{opper,GallaFarmer}, we perform a generating functional analysis of the Sato-Crutchfield dynamics (i.e., the continuous limit of EWA).  This will lead to an effective dynamics that is representative of the continuous limit of the EWA system, for large values of $N$, for typical realizations of the payoffs, and after averaging over the ensemble of random games.  The fixed points of the effective dynamics are far easier to study analytically than those of the the Sato-Crutchfield equations for any particular random game.

Consider the dynamics
\BE\eql{gfa_cts}
    \frac{\dot{x}^\mu_i(t)}{x^\mu_i(t)} &=& - r^{-1} \ln x^\mu_i(t) +
        \sum_{\mathclap{\otheractions{\mu}{i}}} \pdimpayoffs{\mu}{i} \otherstrategies{\mu}{i}{t}- \rho^\mu(t) + h^\mu_i(t).
\EE
This is identical to the Sato-Crutchfield dynamics~\eqr{ewa_cts}, except that we have added arbitrary functions $h^\mu_i(t)$ to generate response functions---these will later be set to zero.  Recall that the normalization term $\rho^\mu(t)$ is defined such that the $x^\mu_i(t)$ have mean~$1$.

We define a generating functional
\BE
  Z[\psi] &=& \intCi{[x]} \delta(\text{equations of motion})\exp\plr{\i \sum_{\mu, i} \inti{t} x^\mu_i(t) \psi^\mu_i(t)},
\EE
where $\delta(\text{equations of motion})$ is used to mean that the integral is performed over realizations of~\eqr{gfa_cts}.  Writing these delta functions in Fourier form yields

\BE\eql{Za}
  Z[\psi] &=& \intCi{[x, \htx]} \exp\Bigg(\i \sum_{\mu, i} \inti{t} \bigg\{ \htx^\mu_i(t) \bigg(\frac{\dot{x}^\mu_i(t)}{x^\mu_i(t)} + r^{-1} \ln x^\mu_i(t)
            - \sum_{\mathclap{\otheractions{\mu}{i}}} \pdimpayoffs{\mu}{i} \otherstrategies{\mu}{i}{t} + \rho^\mu(t) - h^\mu_i(t)\bigg) \nonumber \\
            &&~~+ x^\mu_i(t) \psi^\mu_i(t)\bigg\}\Bigg).
\EE

The factor in this expression depending on the payoff elements is
\begin{equation}
  Z_\Pi = \exp\plr{ -\i \!\!\! \sum_{\mu, i_1, \ldots, i_p} \inti{t} \pdimpayoffsalt{\mu}{i}{i_\mu} \htx^\mu_{i_\mu}(t) \otherstrategies{\mu}{i}{t}}.
\end{equation}
Averaging this over the payoff elements gives
\BE
  \overline{Z_\Pi} &=& \prod_{\allactions{i}{p}} \exp\Bigg\{-\frac{1}{2 N^{p-1}} \sum_{\mu} \inti{t} \inti{t'} 
  \Bigg[\htx^\mu_{i_\mu}(t) \htx^\mu_{i_\mu}(t') \Bigg(\otherstrategies{\mu}{i}{t}\Bigg) \Bigg(\otherstrategiesalt{\mu}{i}{t'}{\lambda}\Bigg) \nonumber \\
  &&+ \Gamma \sum_{\nu \neq \mu} \htx^\mu_{i_\mu}(t) \htx^\nu_{i_\nu}(t') \Bigg(\otherstrategies{\mu}{i}{t}\Bigg) \Bigg(\otherstrategiesalt{\nu}{i}{t'}{\lambda}\Bigg) \Bigg] \Bigg\},
\EE
which can be written as
\be \eql{finalZPi}
  \overline{Z_\Pi} = \exp \Bigg\{ -\frac{N}{2} \inti{t} \inti{t'} \sum_\mu \Bigg( L^\mu(t,t') \prod_{\kappa \neq \mu} C^\kappa(t,t') 
                 + \Gamma \sum_{\nu \neq \mu} K^\mu(t,t') K^\nu(t',t) \prod_{\kappa \notin \{\mu,\nu\}} C^\kappa(t,t') \Bigg) \Bigg\},
\ee
where we have introduced the functions
\BE
  C^\mu(t,t') &= &\frac{1}{N} \sum_i x^\mu_i(t) x^\mu_i(t'), \nonumber \\
  K^\mu(t,t') &= &\frac{1}{N} \sum_i x^\mu_i(t) \htx^\mu_i(t'), \nonumber \\
  L^\mu(t,t') &=& \frac{1}{N} \sum_i \htx^\mu_i(t) \htx^\mu_i(t').
\EE

We can use the expression~\eqr{finalZPi} in \eqr{Za}, introducing the functions $C^\mu$, $K^\mu$, and $L^\mu$ into the integral using delta functions, for example

\BE
  1 &=& \intCi{[C^\mu]} \prod_{t,t'} \delta\plr{C^\mu(t,t') - \frac{1}{N} \sum_i x^\mu_i(t) x^\mu_i(t')} \nonumber \\
    &=& \intCi{[C^\mu, \htC^\mu]} \exp\plr{\i N \inti{t} \inti{t'} \htC^\mu(t,t') \plr{C^\mu(t,t') - \frac{1}{N} \sum_i x^\mu_i(t) x^\mu_i(t')}}.
\EE

The generating functional becomes
\begin{equation} \eql{Zb}
  \overline{Z[\psi]} = \intCi{[C, \htC, K, \htK, L, \htL]} \exp(N(\psi + \Phi + \Omega)),
\end{equation}
where
\BE
  \Psi = \i \sum_\mu \inti{t} \inti{t'} \left(\htC^\mu(t,t') C^\mu(t,t')+ \htK^\mu(t,t') K^\mu(t,t') + \htL^\mu(t,t') L^\mu(t,t')\right)
\EE
results from the introduction of $C$, $K$, and $L$ into the integral,
\be
  \Phi = -\frac{1}{2} \sum_\mu \inti{t} \inti{t'} \Bigg( L^\mu(t,t') \prod_{\kappa \neq \mu} C^\kappa(t,t') 
         + \Gamma \sum_{\nu \neq \mu} K^\mu(t,t') K^\nu(t',t) \prod_{\kappa \notin \{\mu,\nu\}} C^\kappa(t,t') \Bigg)
\ee
results from the average over the payoff elements, and

\BE
  \Omega &=& \frac{1}{N} \sum_{\mu,i} \ln \Bigg\{ \intCi{[x^\mu_i,\htx^\mu_i]} p^\mu_{i,0}(x^\mu_i(0)) \exp\plr{\i \inti{t} x^\mu_i(t) \psi^\mu_i(t)}\exp\plr{\i \inti{t} \htx^\mu_i(t) \plr{\frac{\dot{x}^\mu_i(t)}{x^\mu_i(t)} + \frac{1}{r} \ln x^\mu_i(t) + \rho^\mu(t) - h^\mu_i(t)}} \nonumber \\
           &&\times\exp \Bigg[ -\i \inti{t} \inti{t'} \Bigg( \htC^\mu(t,t') x^\mu_i(t) x^\mu_i(t') 
           + \htK^\mu(t,t') x^\mu_i(t) \htx^\mu_i(t') + \htL^\mu(t,t') \htx^\mu_i(t) \htx^\mu_i(t') \Bigg) \Bigg] \Bigg\}
\EE

contains the integral over $x$ and $\htx$.  Here, $p^\mu_{i,0}(\cdot)$ represents the initial distribution of $x^\mu_i$.

In the limit as $N \rightarrow \infty$, the integrals in~\eqr{Zb} can be performed using the saddle-point method.  Extremising the exponent with respect to $C^\mu$, $K^\mu$, and $L^\mu$ gives the relations
\BE
  \i \htC^\mu(t,t') &=& \frac{1}{2} \sum_{\nu \neq \mu} \Bigg( L^\nu(t,t') \prod_{\kappa \notin \Blr{\mu,\nu}} C^\kappa(t,t') +\Gamma \sum_{\kappa \notin \Blr{\mu, \nu}} K^\nu(t,t') K^\kappa(t',t) \prod_{\lambda \notin \Blr{\mu,\nu,\kappa}} C^\lambda(t,t') \Bigg), \nonumber \\
  \i \htK^\mu(t,t') &=& \Gamma \sum_{\nu \neq \mu} K^\nu(t,t') \prod_{\kappa \notin \Blr{\mu,\nu}} C^\kappa(t,t'), \nonumber \\
  \i \htL^\mu(t,t') &= &\frac{1}{2} \prod_{\kappa \neq \mu} C^\kappa(t,t'),
\EE
while extremisation with respect to $\htC^\mu, \htK^\mu$, and $\htL^\mu$ leads to
\BE
  C^\mu(t,t') &= &\lim_{N \rightarrow \infty} \frac{1}{N} \sum_i \mean{x^\mu_i(t) x^\mu_i(t')}_\Omega, \nonumber \\
  K^\mu(t,t') &= &\lim_{N \rightarrow \infty} \frac{1}{N} \sum_i \mean{x^\mu_i(t) \htx^\mu_i(t')}_\Omega, \nonumber \\
  L^\mu(t,t') &= &\lim_{N \rightarrow \infty} \frac{1}{N} \sum_i \mean{\htx^\mu_i(t) \htx^\mu_i(t')}_\Omega,
\EE
where $\mean{\cdot}_\Omega$ represents a mean taken against a measure defined by ~$\Omega$, see for example the Supplemental Material of \cite{GallaFarmer} for details in a similar calculation for $p=2$.

It can also be seen, from the definition of the generating functional, that we have
\BE
  C^\mu(t,t') &= &- \lim_{N \rightarrow \infty} \frac{1}{N} \sum_i \eval{\madiff{\overline{Z[\psi]}}{\psi^\mu_i(t)}{\psi^\mu_i(t')}}{\psi = h = 0}, \nonumber \\
  K^\mu(t,t') &=& - \lim_{N \rightarrow \infty} \frac{1}{N} \sum_i \eval{\madiff{\overline{Z[\psi]}}{\psi^\mu_i(t)}{h^\mu_i(t')}}{\psi = h = 0}, \nonumber \\
  L^\mu(t,t') &=& - \lim_{N \rightarrow \infty} \frac{1}{N} \sum_i \eval{\madiff{\overline{Z[\psi]}}{h^\mu_i(t)}{h^\mu_i(t')}}{\psi = h = 0}.
\EE
Because of normalization, $Z[\psi=0]=1$ for all $h$, so $L^\mu(t,t')=0$ $\forall t,t'$.  Due to causality, we have $K^\mu(t,t') = 0$ for $t' > t$, so that $K^\mu(t,t') K^\nu(t',t) = 0$.

This leaves $\Psi + \Phi = 0$, and if we set $\psi = 0$, and assume identical perturbations $h^\mu_i(t) = h(t)$ and initial distributions $p^\mu_{i,0}(x) = p_0(x)$ for all players and strategy components, then we have

\BE
  \Omega &=& p \ln \Bigg\{ \intCi{[x,\htx]} p_0(x(0)) 
           \exp\plr{\i \inti{t} \htx(t) \plr{\frac{\dot{x}(t)}{x(t)} + \frac{1}{r} \ln x(t) + \rho(t) - h(t)}} \nonumber \\
           &&\times \exp \Bigg[ - \inti{t} \inti{t'} \Bigg(\Gamma (p-1) K(t,t') C(t,t')^{p-2} x(t) \htx(t') + \frac{1}{2} C(t,t')^{p-1} \htx(t) \htx(t') \Bigg) \Bigg] \Bigg\}
\EE

where we have dropped the distinction between different players and strategy components.  Each degree of freedom then has an effective generating functional

\BE
  Z_\eff &=& \intCi{[x,\htx]} p_0(x(0)) \exp\plr{\i \inti{t} \htx(t) \plr{\frac{\dot{x}(t)}{x(t)} + \frac{1}{r} \ln x(t) + \rho(t) - h(t)}} \nonumber \\
         &&\times  \exp \Bigg[ - \inti{t} \inti{t'} \Bigg(\Gamma K(t,t') C(t,t')^{p-2} x(t) \htx(t') + \frac{1}{2} C(t,t')^{p-1} \htx(t) \htx(t') \Bigg) \Bigg].
\EE

Defining $G(t,t') = - \i K(t,t')$, we have

\BE
  Z_\eff &=& \intCi{[x,\htx]} p_0(x(0)) \exp\plr{\i \inti{t} \htx(t) \plr{\frac{\dot{x}(t)}{x(t)} + \frac{1}{r} \ln x(t) + \rho(t) - h(t)}} \nonumber \\
      &&\times    \exp \Bigg[ - \inti{t} \inti{t'} \Bigg( \i \Gamma G(t,t') C(t,t')^{p-2} x(t) \htx(t') + \frac{1}{2} C(t,t')^{p-1} \htx(t) \htx(t') \Bigg) \Bigg],
\EE

which is identical to the generating functional of the effective dynamics
\BE \eql{effdynh}
  \frac{\dot{x}(t)}{x(t)} &= &\Gamma \inti{t'} G(t,t') C(t,t')^{p-2} x(t') - \frac{1}{r} \ln x(t) - \rho(t) + \eta(t) + h(t),
\EE
where $\eta(t)$ is a Gaussian random variable satisfying $\mean{\eta(t) \eta(t')}_* = C(t,t')^{p-1}$ and $\mean{\eta(t)}_* = 1$, and the functions $G$ and $C$ are determined by
\BE
  G(t,t') &= &\mean{\adiff{x(t)}{h(t')}}_*, \nonumber \\
  C(t,t') &= &\mean{x(t)x(t')}_*,
\EE
with $\mean{\cdot}_*$ used to denote an average over the effective dynamics~\eqr{effdynh}.  Finally setting $h$ to zero, the effective system is
\begin{equation} \eql{effdyn}
  \frac{\dot{x}(t)}{x(t)} = \Gamma \inti{t'} G(t,t') C(t,t')^{p-2} x(t') - \frac{1}{r} \ln x(t) - \rho(t) + \eta(t).
\end{equation}
with $G$, $C$, and $\eta$ defined as above.

\section{Onset of instability in the large-\texorpdfstring{$p$}{p} limit} \secl{straightlines}
Writing $n = p-1$ for convenience, the boundary of the stable region is given by the solution of the following equations:
\begin{align}
  \frac{1}{r} \ln{x} - \Gamma q^{n-1} \chi x - q^{n/2} z + \rho &= 0, \nonumber \\
  \intDlu{z}{-\infty}{\infty}{\pdiff{x}{z}} &= q^{n/2} \chi, \nonumber \\
  \intDlu{z}{-\infty}{\infty}{x^2} &= q \nonumber\\
  \intDlu{z}{-\infty}{\infty}{x} &= 1 \nonumber \\
  \intDlu{z}{-\infty}{\infty}{\plr{\pdiff{x}{z}}^2} &= \frac{q}{n}, \label{eq:stabeqn}
\end{align}
where $\DD z$ is a shorthand for the standard Gaussian measure $\DD z = \frac{\dd z}{\sqrt{2 \pi}} \e^{-z^2/2}$.

For $\Gamma=0$ the order parameters at the boundary of the stable region are given by Eq. (\ref{eq:stabgamma0}). As an ansatz for the region with $\Gamma<0$ we assume that the order parameters and the value of $r$  on the phase boundary  scale with $n$ in the same way as they do for $\Gamma = 0$. We can write 
\BE
  q &= &1 + n^{-1} q', \nonumber \\
  \chi &=& n^{-\half} \chi', \nonumber \\
  r &= &n^{-\half} r', \nonumber \\
  \rho &= &n^{-\half} \rho',
\EE
where all primed variables are of order ${\cal O}(n^0)$.

If we also write $x = 1 + n^{-\half} x'$, and retain only leading-order terms in $q'$, $\chi'$, $r'$, and $\rho'$ t we obtain from Eq. (\ref{eq:stabeqn}): 
\begin{align}
  \begin{split}
    \frac{1}{r'} \ln\plr{1 + n^{-\half} x'} - n^{-\half} (1 + q'/2) z + n^{-1} \rho' \\
  - n^{-1} \Gamma (1+q') \chi' - n^{-\frac{3}{2}} \Gamma (1+q') \chi' x' &= 0,
  \end{split} \nonumber \\
  \intDlu{z}{-\infty}{\infty} \pdiff{x'}{z} &= \plr{1 + \frac{q'}{2}} \chi', \nonumber \\
  \intDlu{z}{-\infty}{\infty} x'^2 &= q', \nonumber \\
  \intDlu{z}{-\infty}{\infty} x' &= 0, \nonumber \\
  \intDlu{z}{-\infty}{\infty} \plr{\pdiff{x'}{z}}^2 &= 1+\frac{q'}{n}. \label{eq:pstabeqn}
\end{align}

The linear term in $x'$ in the first of these equations is dominated by the log term except at large x.  Specifically, by using the approximation $W_{-1}(y) \approx \ln(-y)$ as $y \rightarrow \infty$, it can be seen that the linear term reaches the size of the log term when the value of $x'$, to leading order, is
\begin{equation}
  x' \approx x_l = \frac{n^{\frac{3}{2}} \ln{n}}{- \Gamma r' (1+q') \chi'},
\end{equation}
while to leading order $z$ is
\begin{equation}
  z \approx z_l = \frac{n^\half \ln n}{r'\plr{1+\frac{q'}{2}}}
\end{equation}

It remains only to show that the region of the real line beyond~$z_l$ makes a vanishing contribution to the integrals in Eqs. (\ref{eq:pstabeqn}).  By ignoring the linear term for~$z < z_l$, and the log term for~$z > z_l$, the integrals over these two regions can be approximated analytically.  In each case, the~$z > z_l$ contribution shrinks more quickly as~$n$ grows.

Neglecting the linear term in the first relation in Eq. (\ref{eq:pstabeqn}) altogether is equivalent to making the approximation~$x=1$ in the linear term in the first equation of (\ref{eq:stabeqn}).  This yields a system of the same form as the~$\Gamma=0$ special case, except for an additional constant term, which can be solved exactly in the same manner.  So, to leading order, the parameters $r$, $\chi$, and $q$ take constant values along the stability curve for large $p$, while $\rho$ takes the value
\begin{equation}
  \rho = \plr{\half + \Gamma} \sqrt{\frac{\e}{n}}.
\end{equation}
This value for~$\rho$ scales with~$n$ in the same way as it does in our ansatz, so the ansatz is indeed valid for all negative values of~$\Gamma$. This demonstrates that $r = \sqrt{\e (p-1)}$ is a solution of the equations for the onset of instability in the limit of large $p$.

\section{Heuristic classification of the dynamic behavior} \secl{numerics}
It is not necessarily straightforward to classify the long-term behavior of even low-dimensional dynamical systems using empirical data. For high-dimensional systems such as the EWA dynamics for games with large number of players and/or strategies this task can be extremely challenging.

We experience two major difficulties.  Firstly, in some regions of parameter space, transient behavior can last for a very long time, and can appear chaotic for all intents and purposes even though the system eventually reaches a stable fixed point.  Secondly, characterising chaos using Lyapunov exponents or measures of dimension can be problematic for such large systems. the Jacobians of the system for example can be badly conditioned. 

However, these difficult cases are not the norm, and we can use heuristics to classify behavior as convergence to a stable fixed point, convergence to a limit cycle, or chaos, with a high degree of accuracy.

For a given set of parameters and initial conditions, we iterate the EWA system for a maximum of $500 000$ time steps, split into batches of $10 000$ steps.  After each batch, we explicitly check for the appearance of fixed points or limit cycles.  If the relative difference between the maximum and minimum values of each strategy component was less than $1\%$, we assume a stable fixed point has been found.  If there is a~$\tau$ such that all of~$x^\mu_i(t+\tau)$, $x^\mu_i(t+2\tau)$, etc. (where $t$ was the time at the start of the batch) have components within~$0.1\%$ of the components of~$x^\mu_i(t)$, then we assume a stable limit cycle has been found.  Otherwise, we continue to the next batch.  If convergence has not been detected after $500 000$ time steps, we assume the system is chaotic.

This heuristic was used to produce the plots in Figs. \figr{fp_heat_maps} and in Appendix \secr{numapp}.
\\ ~ \\

\section{Further numerical results: limit cycles and multiplicity of fixed points}\secl{numapp}
\subsection{Competitive games ($\Gamma<0$)}
In Fig. \figr{lc_heat_maps} we show the likelihood of converging to a limit cycle  for games with negatively correlated payoff matrix elements, i.e. games in which players compete agains each other ($\Gamma<0$).  For intermediate values of $\alpha$, just smaller than those for which stable fixed points are ubiquitous, limit cycles are seen very commonly.  However, at small values of $\alpha$, fixed points or limit cycles are achieved only rarely---chaos is the norm.

\begin{figure}
\includegraphics[scale=0.5]{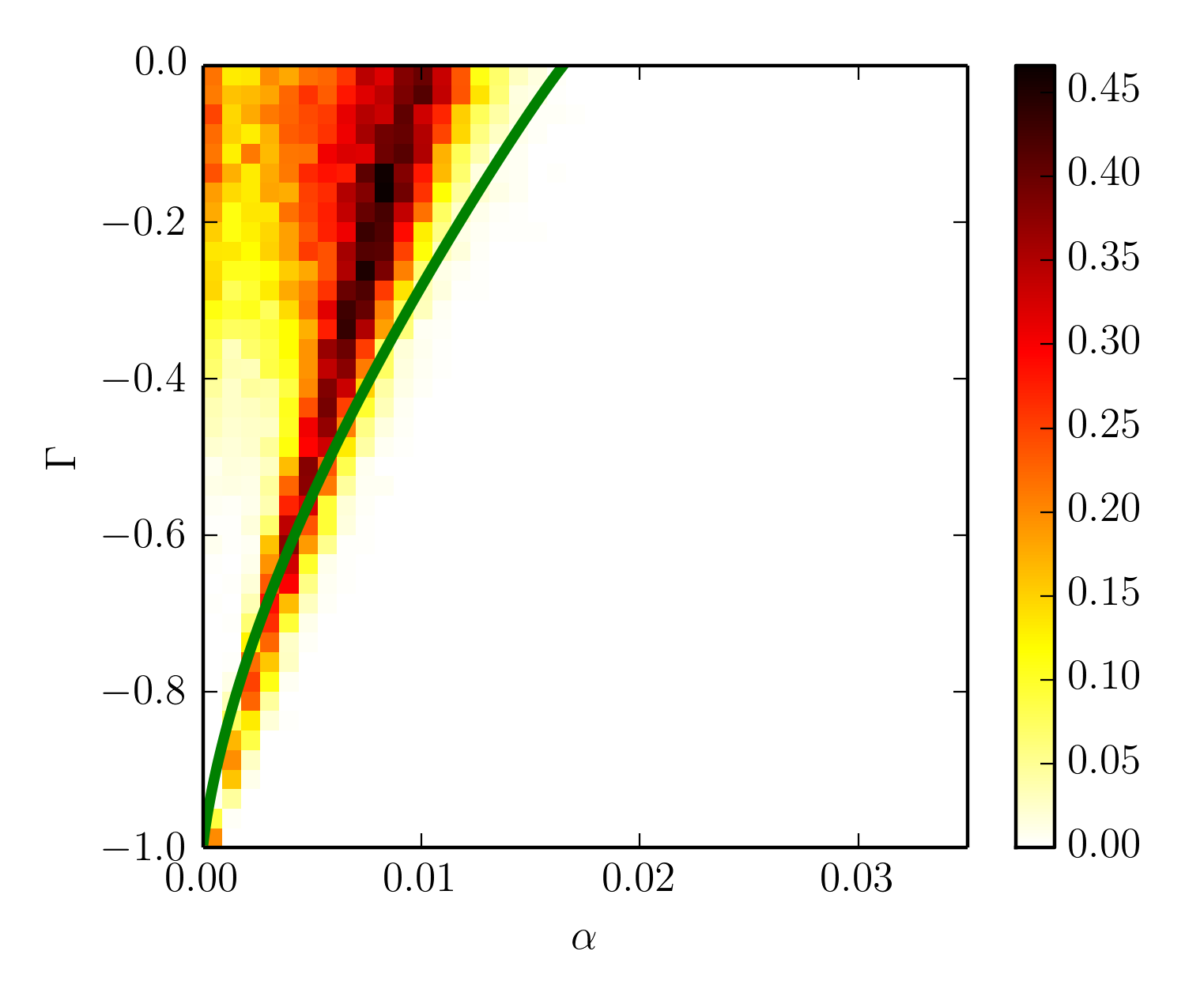}
\includegraphics[scale=0.5]{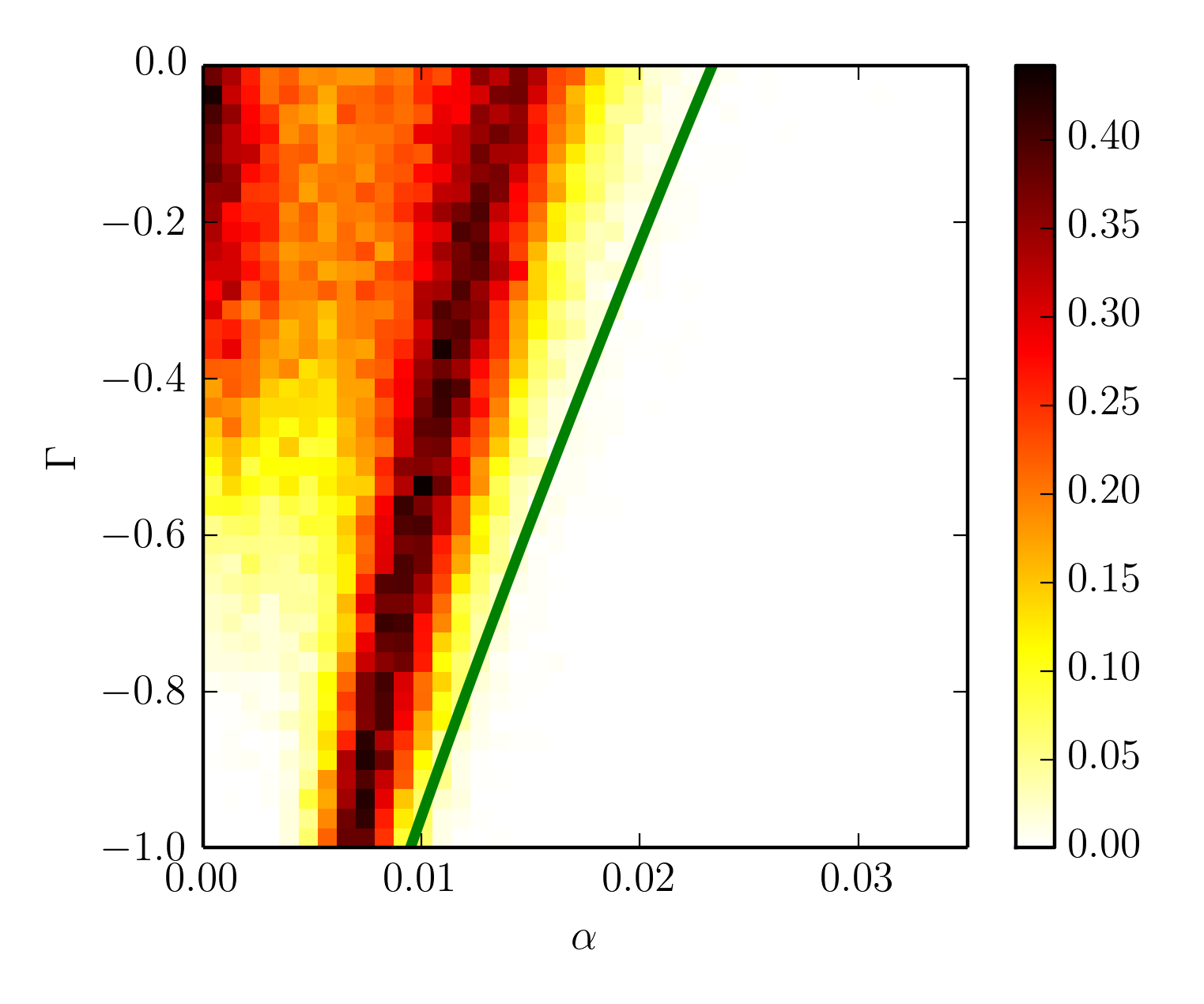}
\\
$p=2$~~~~~~~~~~~~~~~~~~~~~~~~~~~~~~~~~~~~~~~~~~~~~~~~~~~ $p=3$
\\ ~\\
\includegraphics[scale=0.5]{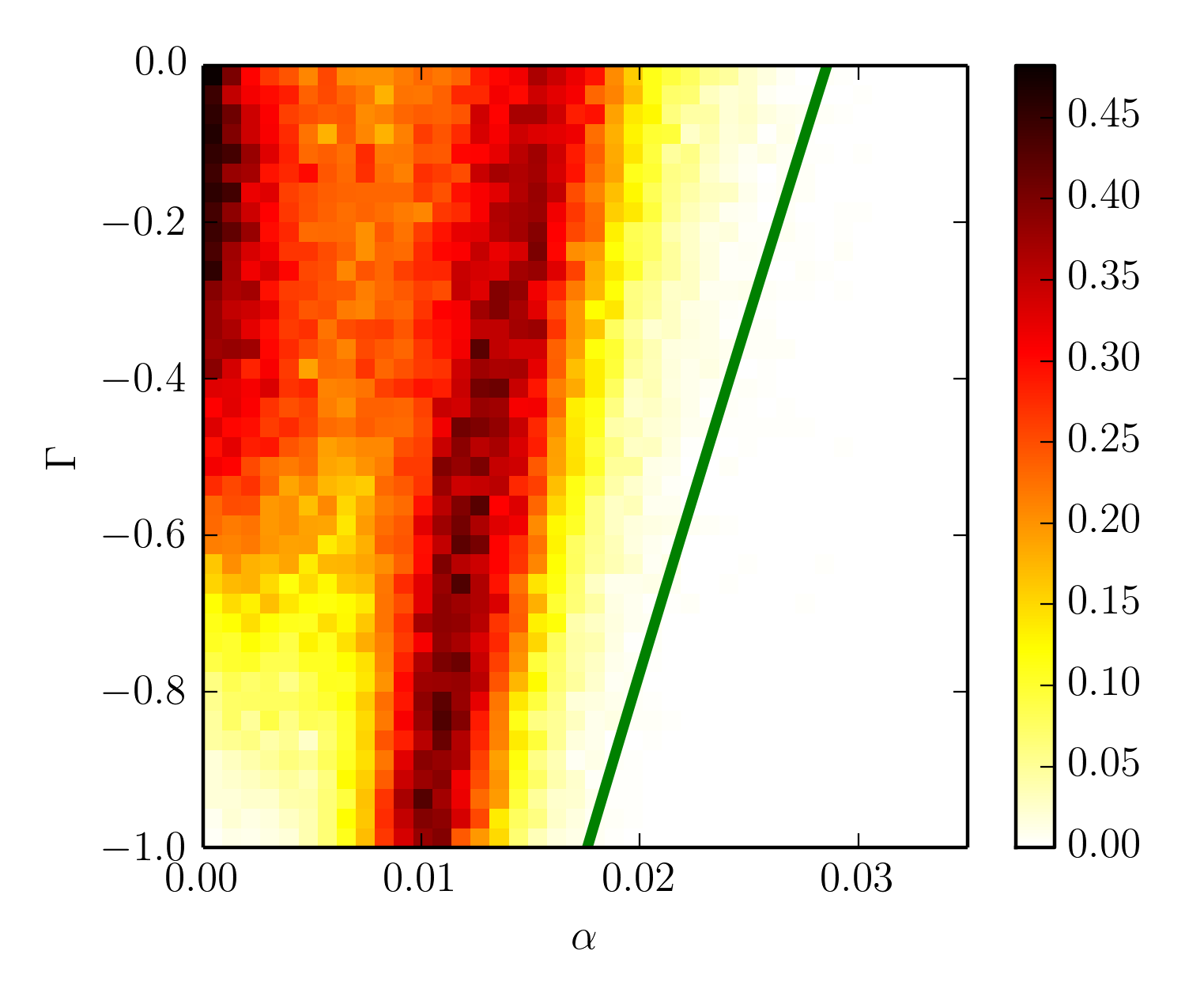}
\includegraphics[scale=0.5]{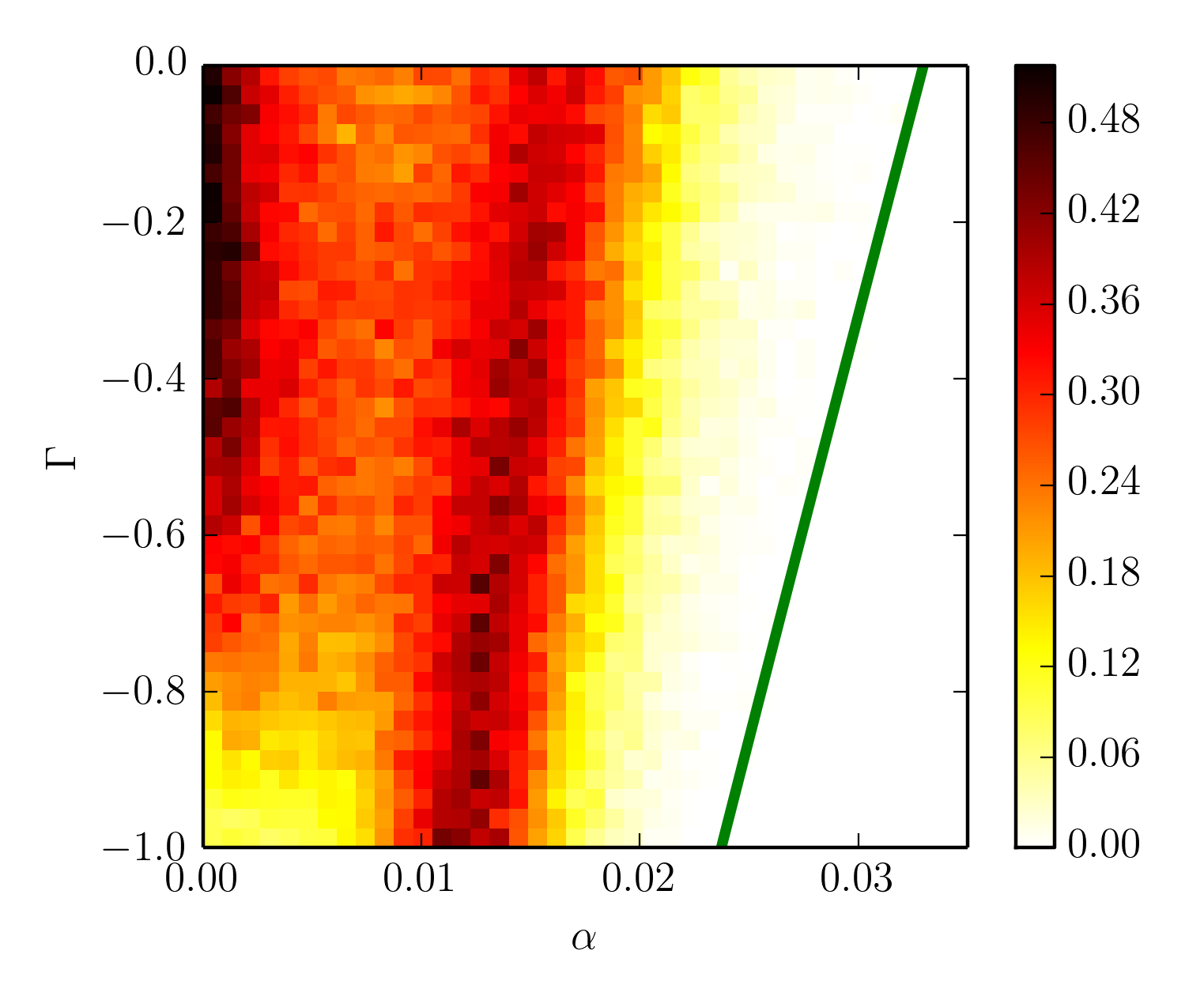}
\\
$p=4$~~~~~~~~~~~~~~~~~~~~~~~~~~~~~~~~~~~~~~~~~~~~~~~~~~~ $p=5$
 \caption{Heat maps showing the fraction of $500$ random initial conditions for which the EWA system converged to a limit cycle according to the heuristic in Appendix~\secr{numerics}, for negative~$\Gamma$.  Limit cycles appear in a narrow band at intermediate values of $\alpha$.  The green curves show the boundaries of the stable region as derived in section~\secr{stabcurves}. }
  \figl{lc_heat_maps}
\end{figure}

\subsection{Positively correlated payoffs ($\Gamma>0$)}

For positive values of the competition parameter, chaotic dynamics is rarely observed (though chaotic-appearing transients are frequently seen).  Instead, for smaller values of~$\alpha$ and~$\Gamma$, limit cycles are very common, as shown in Fig. \figr{pos_lc_heat_maps}.  In the rest of this region, EWA consistently converges to a fixed point.  However, for small values of~$\alpha$ and large values of~$\Gamma$, there are many distinct fixed points that the dynamics can converge to for a given payoff matrix. This is shown in Fig. \figr{pos_multiplicity_heat_maps}.

\begin{figure}[h!!!]
\includegraphics[scale=0.5]{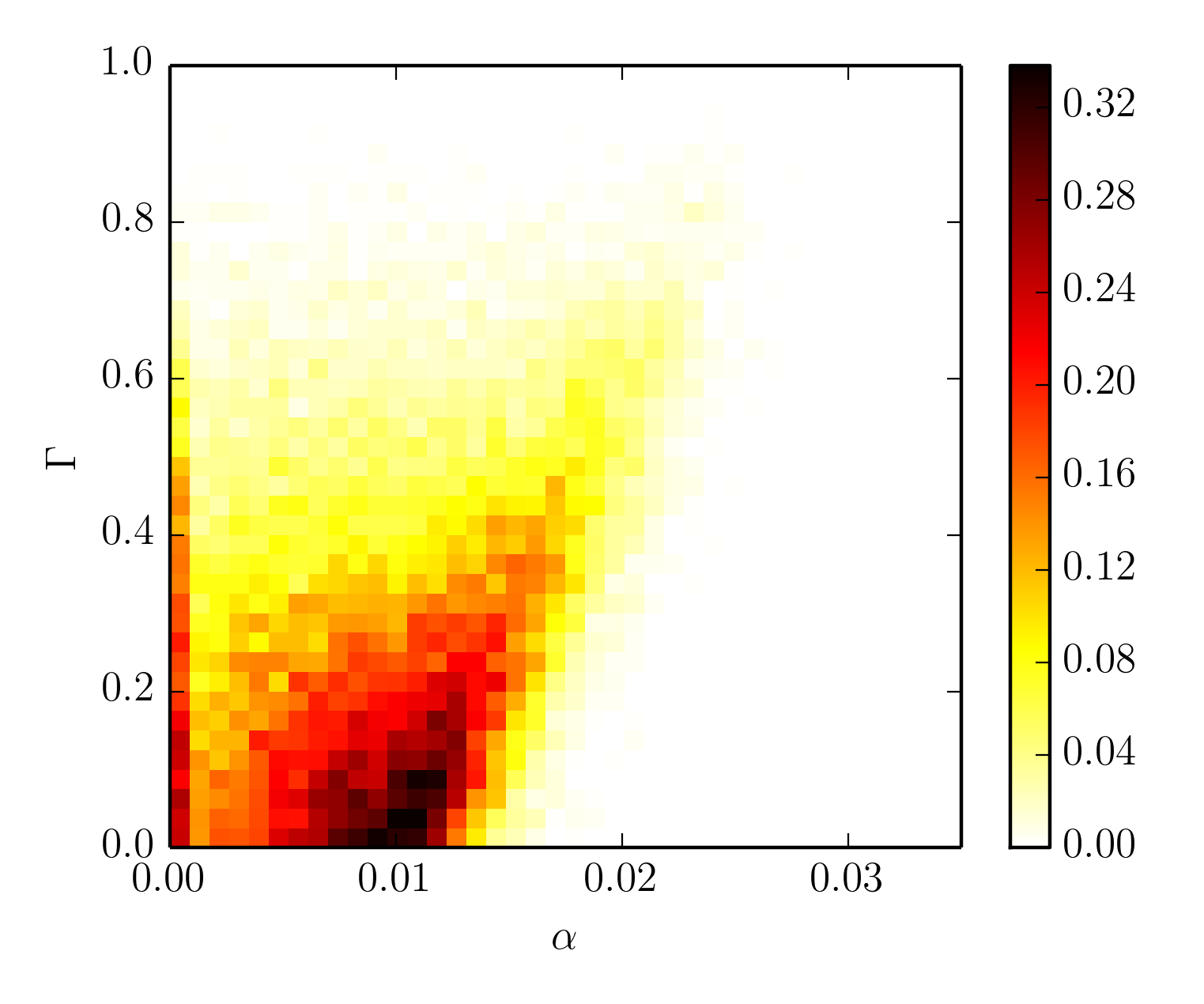}
\includegraphics[scale=0.5]{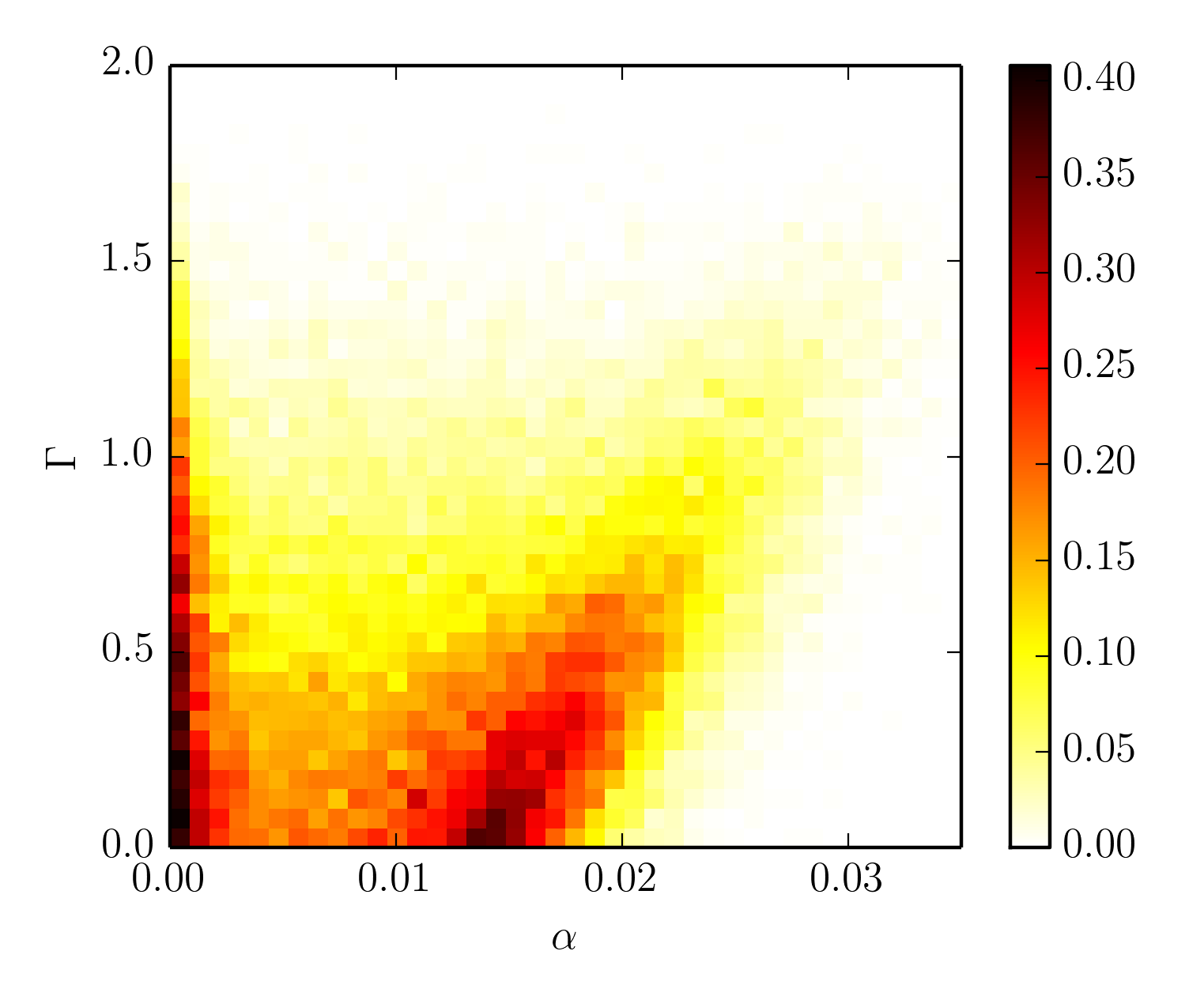}
\\
$p=2, N=50$ ~~~~~~~~~~~~~~~~~~~~~~~~~~~~~~~~~~~~~~~~$p=3, N=12$
\\ ~ \\
\includegraphics[scale=0.5]{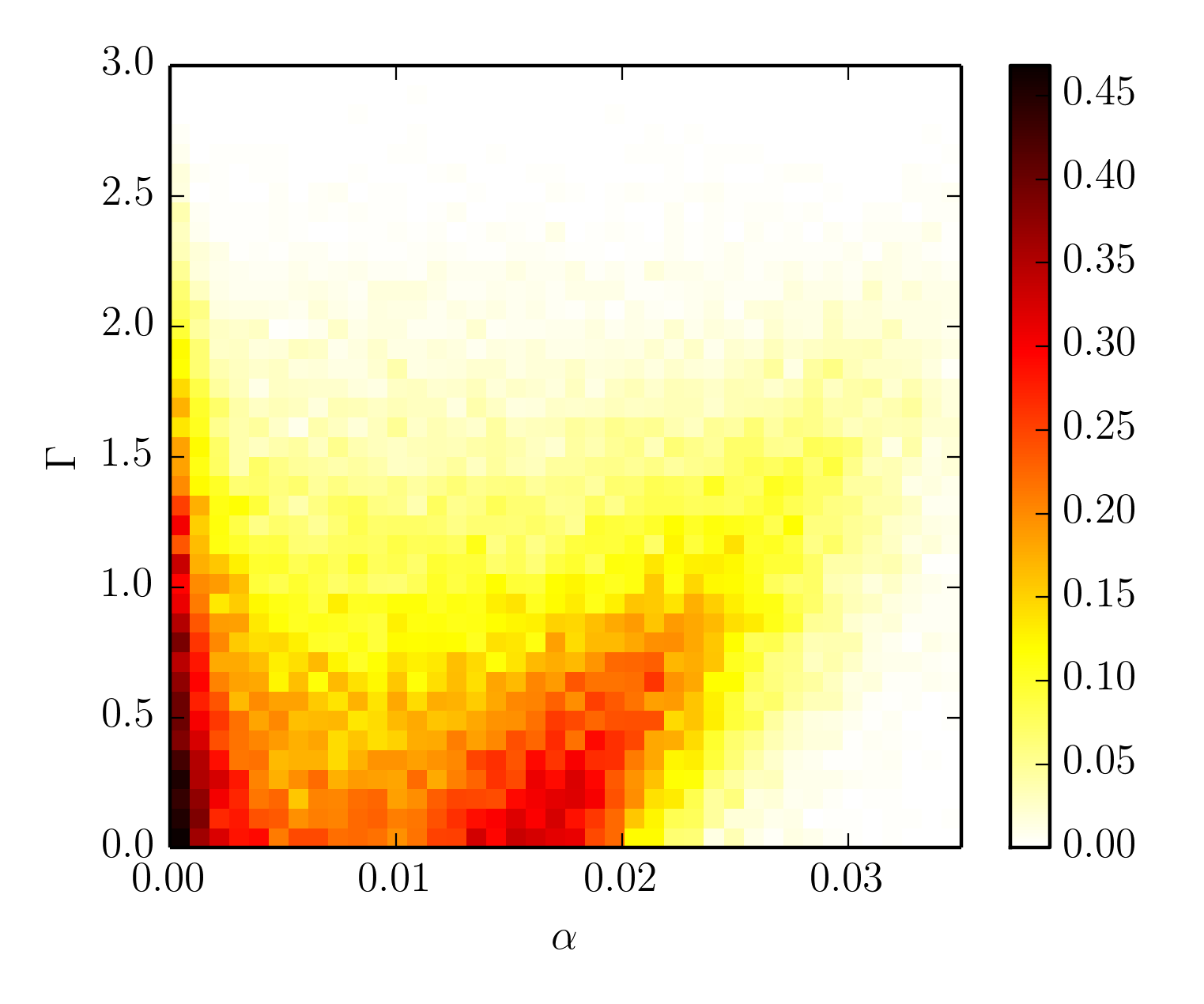}
\includegraphics[scale=0.5]{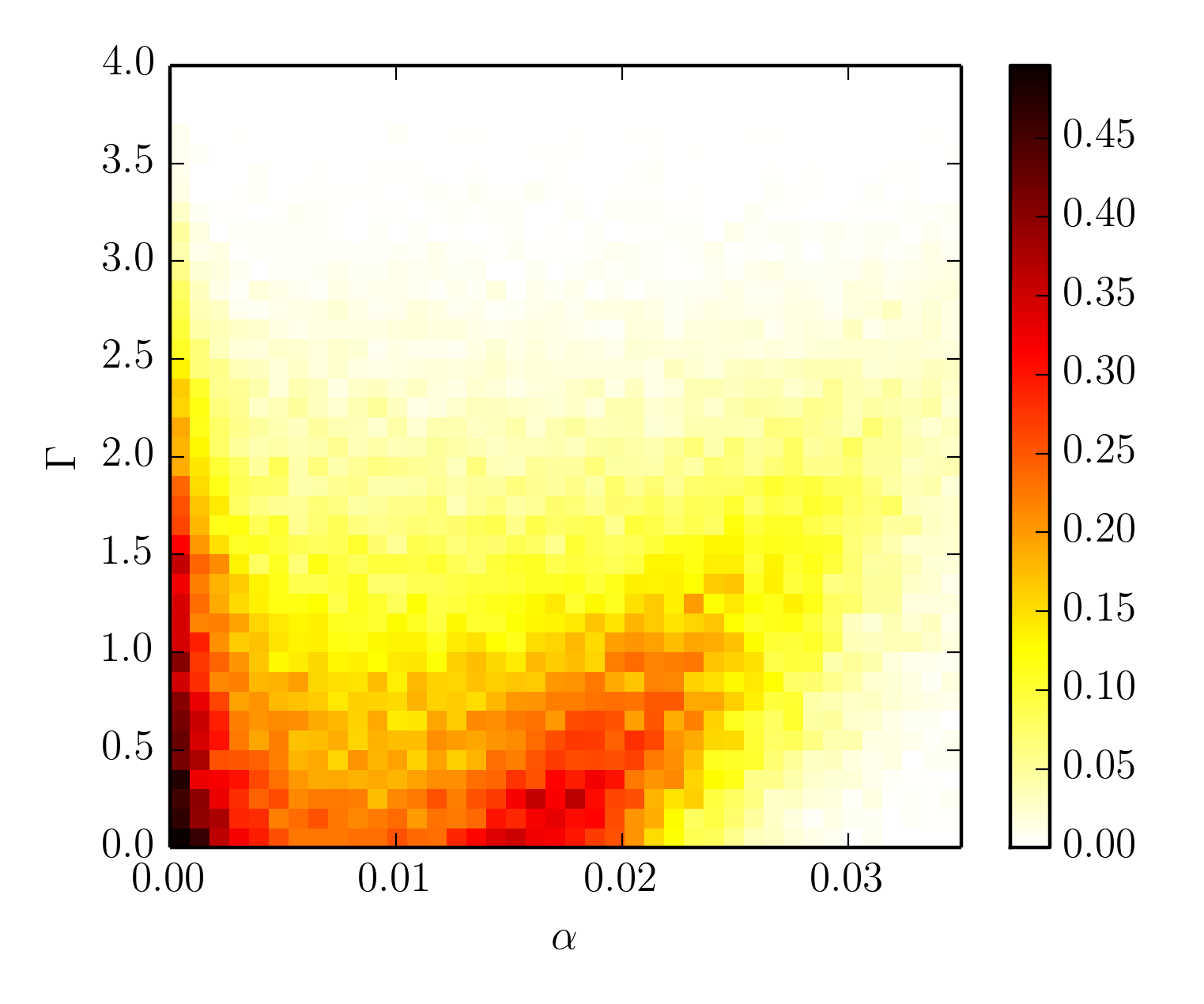}
\\
$p=4, N=6$ ~~~~~~~~~~~~~~~~~~~~~~~~~~~~~~~~~~~~~~~~~~$p=5, N=4$
\caption{Heat maps showing the fraction of $500$ random initial conditions for which the EWA system converged to a limit cycle according to the heuristic in Appendix~\secr{numerics}, for positive~$\Gamma$.  Limit cycles appear most commonly when the payoffs are weakly correlated.}
  \figl{pos_lc_heat_maps}
\end{figure}

\begin{figure}[h!!!]
  \centering
 \includegraphics[scale=0.5]{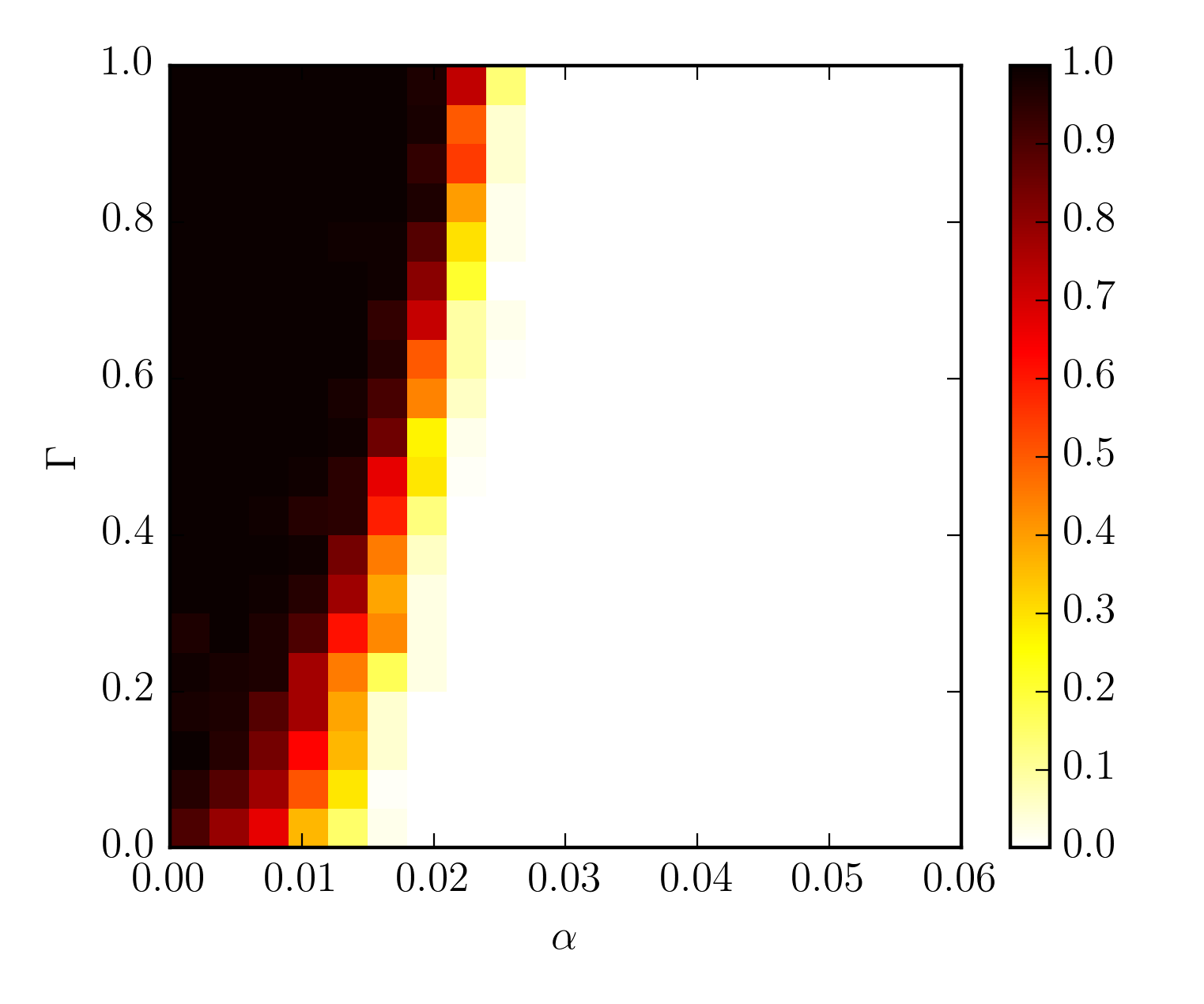}
\includegraphics[scale=0.5]{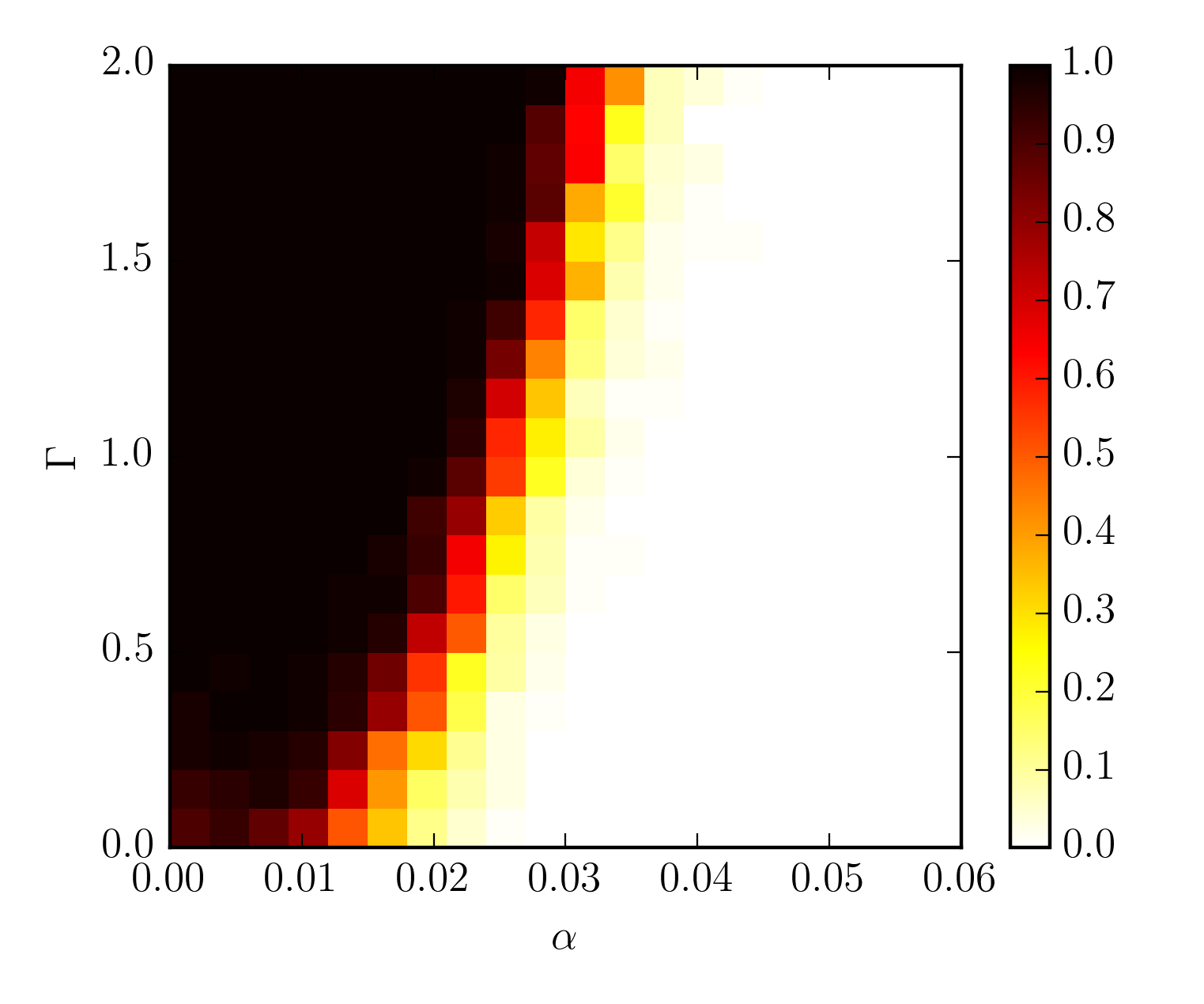}
\\
$p=2, N=50$ ~~~~~~~~~~~~~~~~~~~~~~~~~~~~~~~~~~~~~~~~$p=3, N=12$
\\ ~ \\
 \includegraphics[scale=0.5]{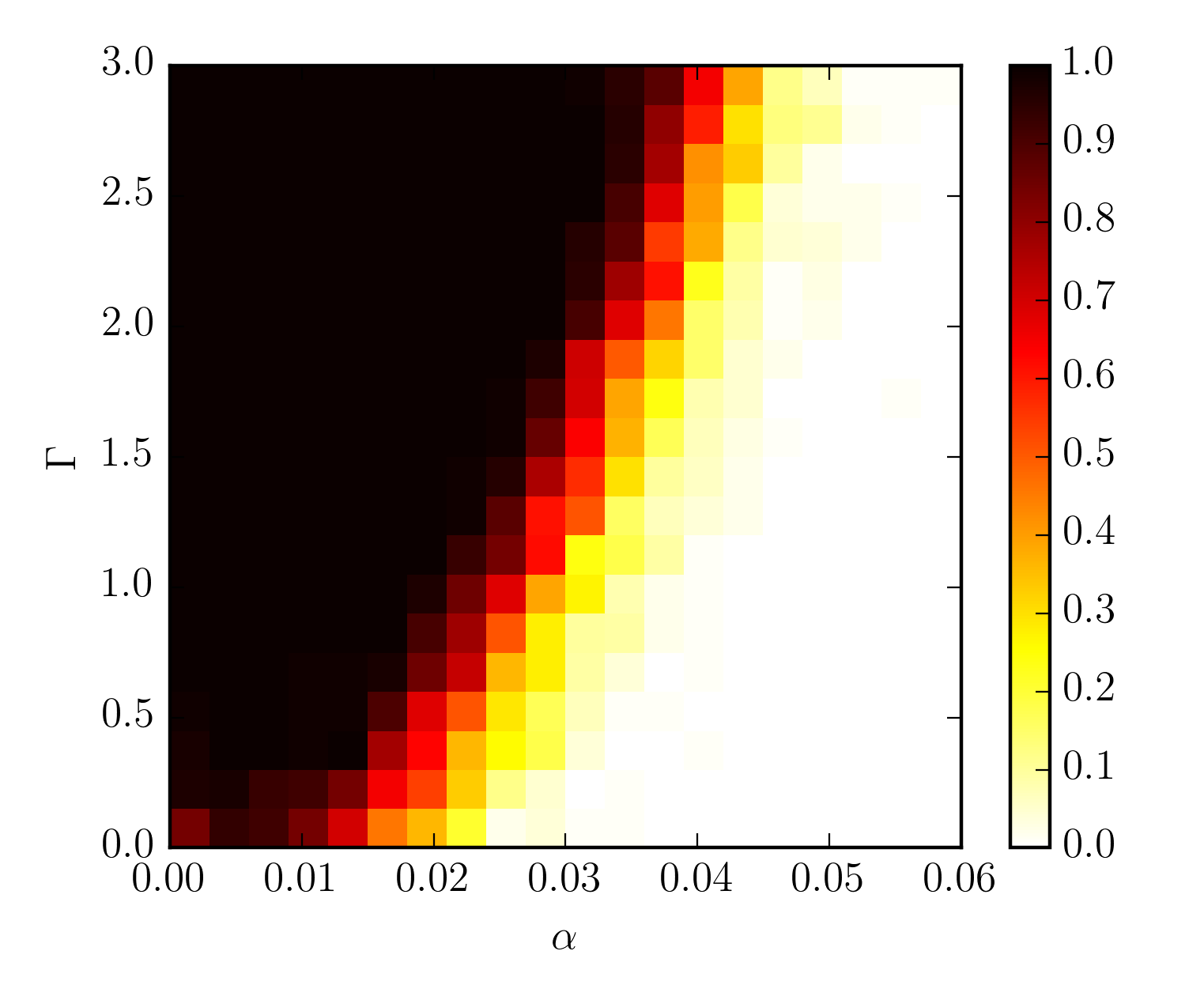}
\includegraphics[scale=0.5]{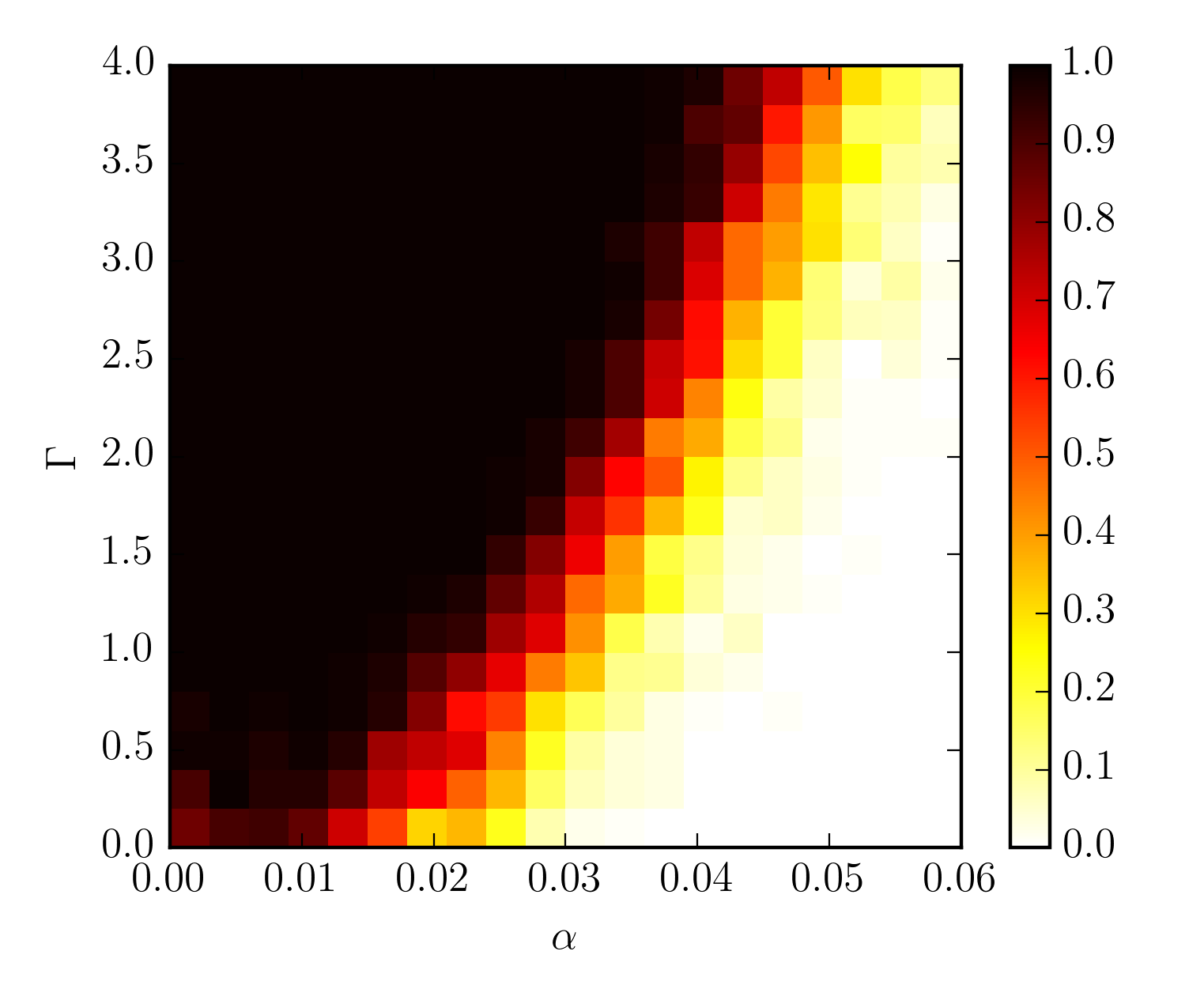}
\\
$p=4, N=6$ ~~~~~~~~~~~~~~~~~~~~~~~~~~~~~~~~~~~~~~~~~~$p=5, N=4$

  \caption{Heat maps showing the fraction of 20 independent payoff matrices for which the EWA dynamics converged to multiple distinct fixed points for different initial conditions.  For each payoff matrix, the EWA system was iterated for 100 different initial conditions, with fixed points being detected as explained in Appendix \secr{numerics}.  Fixed points were considered to be identical if the relative distance between each component was less than $0.1$. }
  \figl{pos_multiplicity_heat_maps}
\end{figure}

\begin{figure}
\includegraphics[scale=0.5]{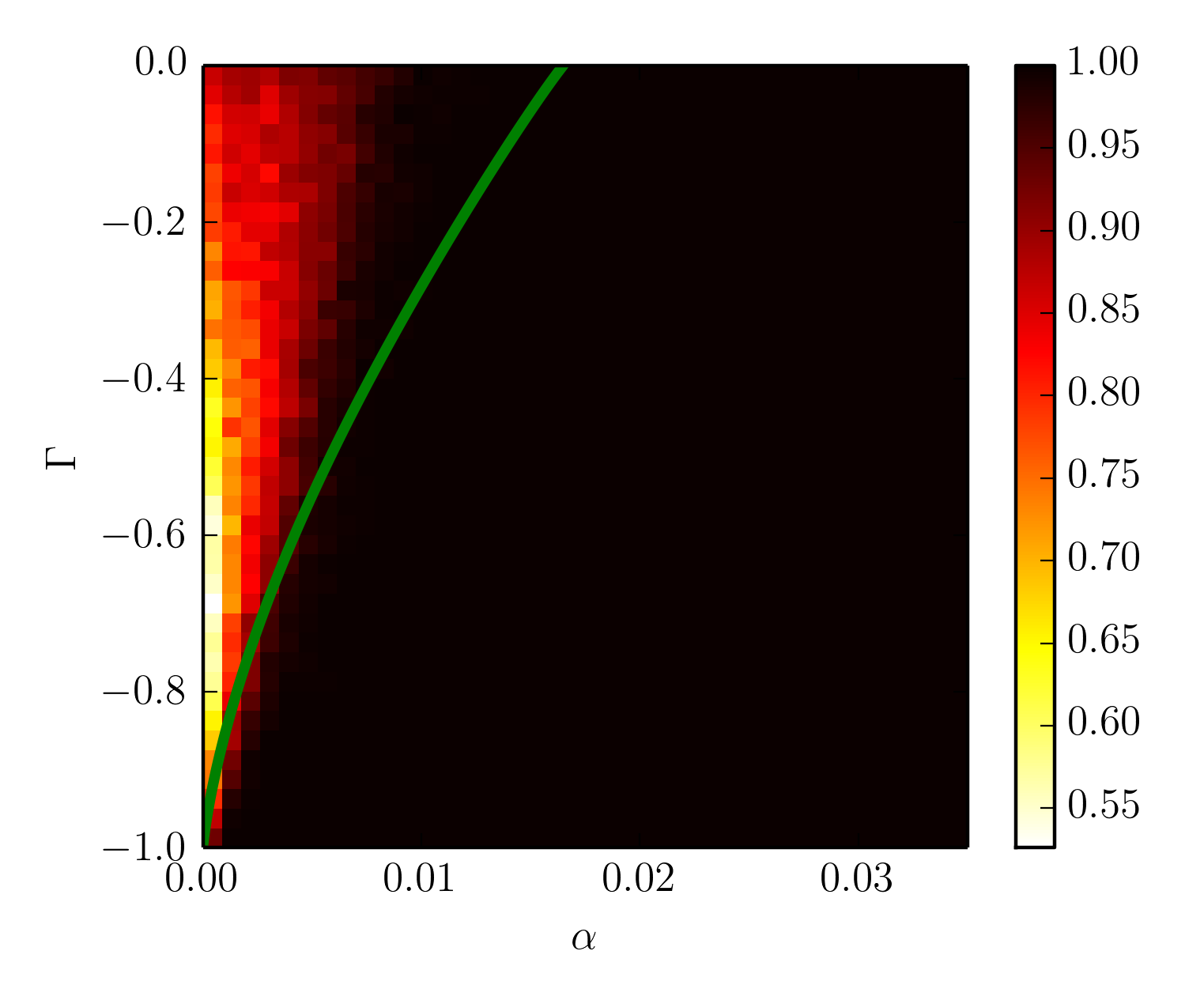}
\\
(a) $p=2, N=10$.
\\    
\includegraphics[scale=0.5]{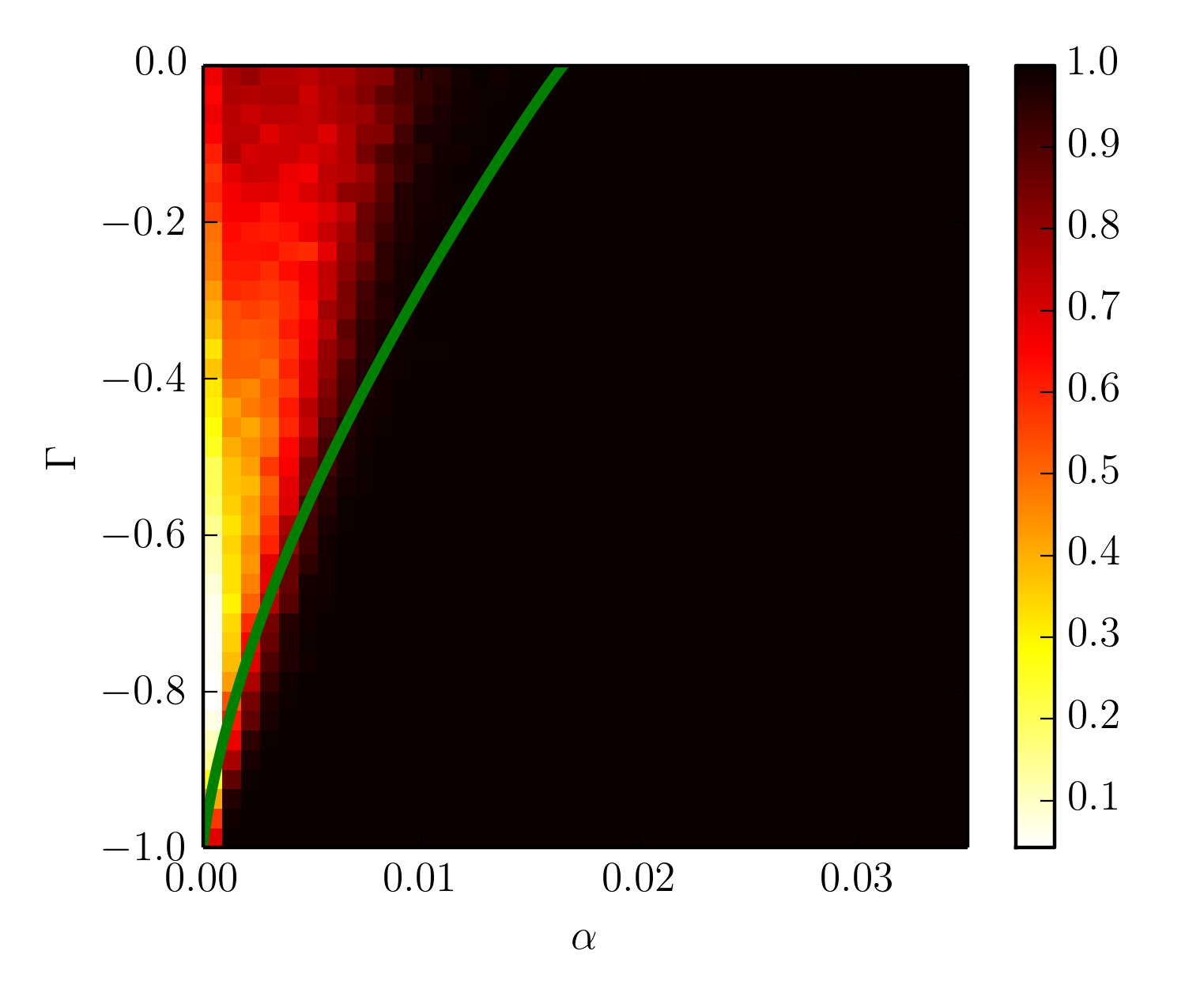}
\\
(b) $p=2, N=20$.
\\
 \includegraphics[scale=0.5]{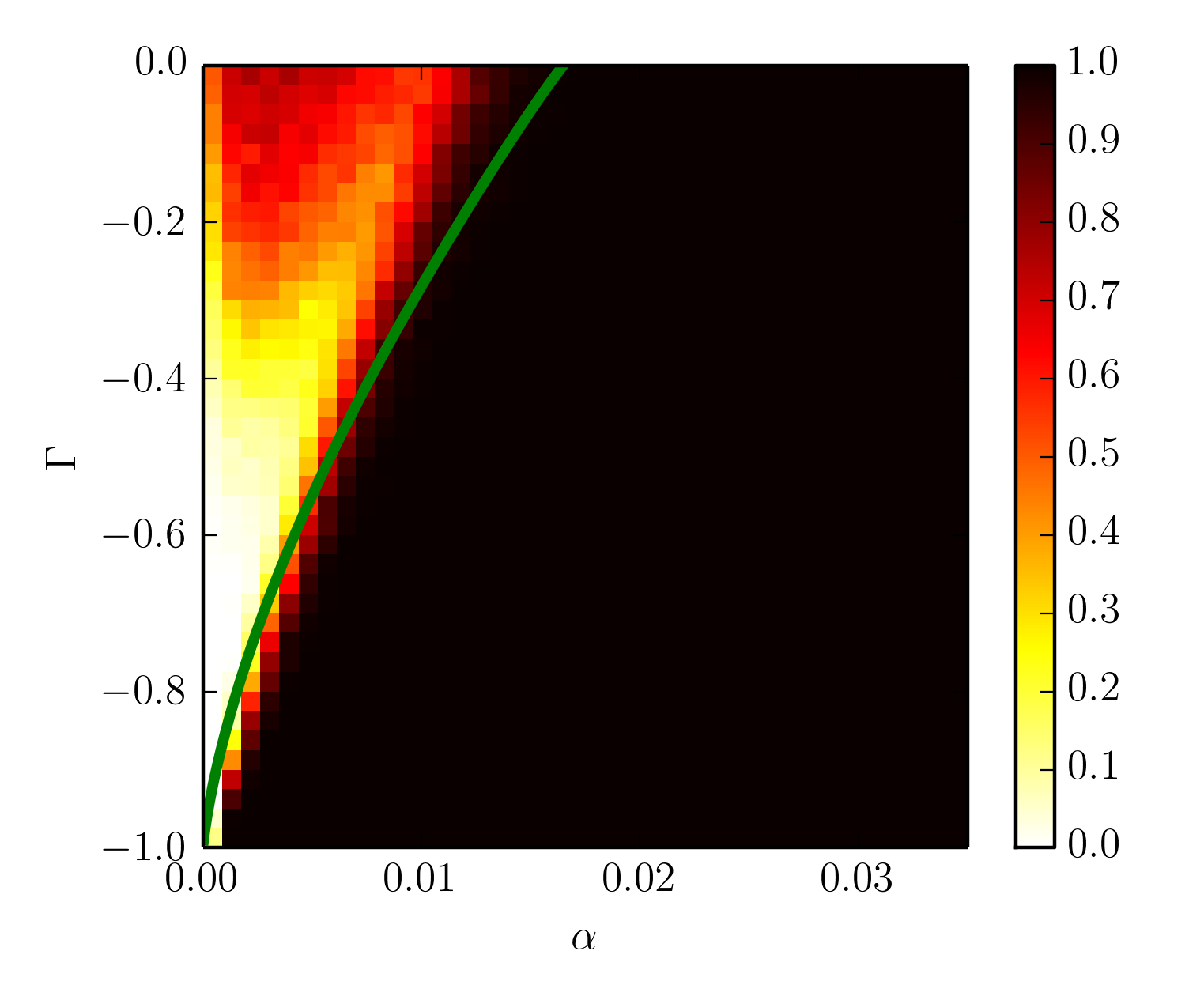}
\\
$p=2, N=50$
\\
    \includegraphics[scale=0.5]{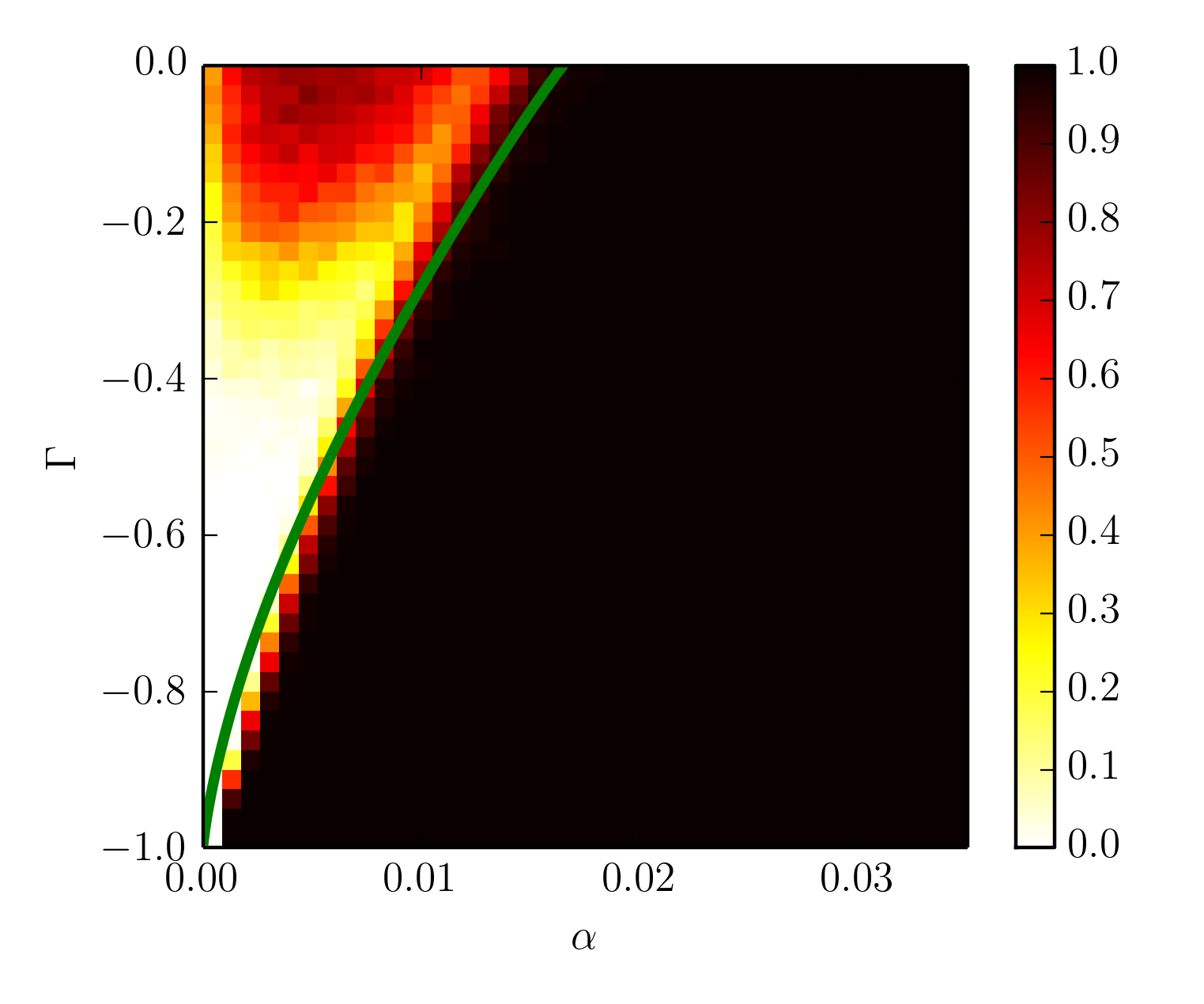}
\\
$p=2, N=100$
\\
\caption{Heat maps showing the dependence on $N$ of the stable region for $p=2$ and $\Gamma<0$.  For each set of parameters the system was iterated for $500$ random initial conditions. The heat maps show the fraction that converged to a fixed point (numerical convergence criteria are described in Appendix \secr{numerics}).  The size of the unstable region grows with the size $N$ of the payoff matrix, but begins to converge around $N=50$.  The green curves are the stability curves as derived in section~\secr{stabcurves}.}
  \figl{N_heat_maps}
\end{figure}


\begin{thebibliography}{99}
\bibitem{f1} Note that the fixed point reached in the stable regime is only a Nash equilibrium in the limit $\alpha\to 0$. When $\alpha > 0$ the players are effectively assuming their opponent's behavior is non-stationary, and that more recent moves are more useful than moves in the distant past.
\bibitem{f2} In contrast to financial markets, for the behavior we observe here the distribution of heavy tails decay exponentially (as opposed to following a power law).  We hypothesize that this is because the players in financial markets use a variety of different timescales $\alpha$.
\end{thebibliography}
\end{document}